\documentclass[
 aps,
 prx,
 reprint,
 superscriptaddress,
 longbibliography,
 amsmath,
 amssymb
]{revtex4-2}

\usepackage{graphicx}
\usepackage{bm}
\usepackage{booktabs}
\usepackage{multirow}
\usepackage{array}
\usepackage{xcolor}
\usepackage{float}

\graphicspath{{figs/}}

\newcolumntype{L}[1]{>{\raggedright\arraybackslash}p{#1}}

\newcommand{\ket}[1]{\left|#1\right\rangle}
\newcommand{\bra}[1]{\left\langle#1\right|}

\newcommand{\ghzvars}{(z,c,s)}

\begin{document}

\title{Sparsified Kolmogorov--Arnold Networks for Interpretable Quantum State Tomography}

\author{Xinge Wu}
\affiliation{National Supercomputing Center in Zhengzhou, Zhengzhou University, Zhengzhou 450001, China}
\affiliation{School of Computer and Artificial Intelligence, Zhengzhou University, Zhengzhou 450001, China}
\author{Huaxin Wang}
\affiliation{National Supercomputing Center in Zhengzhou, Zhengzhou University, Zhengzhou 450001, China}
\affiliation{School of Computer and Artificial Intelligence, Zhengzhou University, Zhengzhou 450001, China}
\author{Jiajun Liu}
\affiliation{National Supercomputing Center in Zhengzhou, Zhengzhou University, Zhengzhou 450001, China}
\affiliation{School of Computer and Artificial Intelligence, Zhengzhou University, Zhengzhou 450001, China}
\author{Ruiqing He}
\affiliation{School of Communication and Artificial Intelligence, School of Integrated Circuits, Nanjing Institute of Technology, Nanjing 211167, China}
\author{Jiandong Shang}
\affiliation{National Supercomputing Center in Zhengzhou, Zhengzhou University, Zhengzhou 450001, China}
\author{Hengliang Guo}
\affiliation{National Supercomputing Center in Zhengzhou, Zhengzhou University, Zhengzhou 450001, China}
\author{Qiang Chen}
\thanks{Corresponding author: qiangchen@zzu.edu.cn}
\affiliation{National Supercomputing Center in Zhengzhou, Zhengzhou University, Zhengzhou 450001, China}
\date{June 10, 2026}

\begin{abstract}
Machine-learning approaches to quantum state tomography can achieve high reconstruction fidelity, but the physical structure used by the trained model often remains implicit. Here we ask whether a sparsified Kolmogorov--Arnold Network (KAN) can be used not only as a regressor, but also as an inspectable reconstruction rule whose internal organization can be checked against known Pauli structure. We study a controlled three-qubit GHZ-family benchmark in which all 63 non-identity Pauli expectation values are used to reconstruct three GHZ-subspace variables: the population imbalance $z$, the real off-diagonal component $c$, and the imaginary off-diagonal component $s$. Under finite-shot sampling and depolarizing noise, external ablation identifies the extended 12-channel GHZ-relevant Pauli set from the 63 measurements, with exact top-12 recovery across the tested shot counts and depolarizing-noise strengths. These support patterns remain stable across multi-seed random-initialization and noise-level analyses, and collapse under random-label controls. The dominant pruned input-hidden-output pathways organize Z-type population observables and X/Y off-diagonal observables in a pattern consistent with the analytic GHZ Pauli grouping, and sparse formula recovery recovers the canonical signed Pauli relations.  The contribution of the KAN is therefore pathway-level structural interpretability within a neural reconstruction model, rather than superior sparse regression. Together with negative controls, these probes provide a consistency chain for auditing learned reconstruction rules against known physical structure.
\end{abstract}

\maketitle

\section{Introduction}

Quantum state tomography is a key tool for diagnosing prepared quantum states, benchmarking quantum processors, and validating entanglement-generation protocols~\cite{ParisRehacek2004,James2001,LvovskyRaymer2009}. Its difficulty grows rapidly with Hilbert-space dimension: the number of parameters in a generic density matrix increases exponentially with qubit number, while measurement budgets and calibration accuracy remain finite~\cite{Aaronson2007,Gross2010,Cramer2010,Flammia2012,Haah2017}. Measurement-efficient alternatives, including compressed-sensing, permutationally invariant, and shadow-based approaches, have therefore been developed for structured tomography settings~\cite{Toth2010,Liu2012ExpCS,Huang2020Shadows}. Practical reconstruction must contend with finite sampling and estimator-dependent behavior~\cite{Hradil1997,BlumeKohout2010,Huszar2012,Qi2013}. Experimental pipelines also depend on assumptions about state preparation and measurement, which can cause a mismatch between idealized tomography protocols and measured data~\cite{Palmieri2020}. These constraints motivate reconstruction methods that are accurate under noisy measurements and transparent enough to reveal when their assumptions fail.

Machine-learning-based approaches have increasingly been used to learn measurement-to-state maps, reconstruct quantum states from incomplete or noisy data, and improve the efficiency of repeated tomography workflows~\cite{Torlai2018,Carrasquilla2019,Xin2019,Palmieri2020,Lohani2023,Gaikwad2024,Wei2024NeuralShadow,MacaronePalmieri2025DeepCS,Hsieh2024Fock,Li2025GenerativeQST,Jabeen2026QMLQST,Krisnanda2025Reservoir}. Recent work has explored feed-forward networks, generative models, neural quantum-state reconstructions, adaptive strategies, attention-based and transformer architectures, convolutional and symmetry-aware models, and experiment-facing reconstruction pipelines~\cite{Quek2021,Koutny2022,Cha2022,Palmieri2024Attention,An2024Unified,Ma2025QAT,Maragnano2025Permutation,Schmale2022,Ma2024,Morgillo2024}. These works show that neural models can be effective reconstruction tools, especially in restricted or structured settings. Related neural-state and machine-learning approaches in quantum many-body physics further show that neural representations can encode structured quantum data compactly~\cite{Carleo2017,Carrasquilla2017}. At the same time, recent sample-complexity analyses indicate that the advantages and boundaries of neural reconstruction should be assessed systematically rather than inferred from accuracy alone~\cite{Zhao2024SampleComplexity,Anshu2024Complexity}. High reconstruction fidelity alone does not reveal which measurement observables the model uses. Nor does it show whether the learned representation follows the physical structure of the target family. A trained network may rely on correlations that are difficult to relate to the underlying measurement physics, redundant channels, column-order artifacts, or dataset-specific shortcuts. This motivates a model-inspection perspective focused on Pauli-resolved input selection, internal routing, and formula-level consistency~\cite{Rudin2019,Murdoch2019,Roscher2020}.

Kolmogorov--Arnold Networks (KANs) provide a practical architecture for this inspection problem~\cite{Liu2024KAN,Liu2024KAN2,Somvanshi2025Survey}. In a KAN, learnable one-dimensional functions are placed on edges. After pruning, this edge-based structure allows sparse routing to be inspected and input-hidden-output pathways can be ranked. Dense MLP layers do not naturally define named input--hidden--output routes. In a KAN, by contrast, trained edges can be inspected, pruned, and traced as routes from Pauli inputs to output variables. KANs have recently been applied to quantum architecture search, model discovery, and scientific or physical modeling~\cite{Kundu2024KANQAS,Liu2024KAN2,Koenig2024KANODE,Wu2024KANPhysics,Panahi2025KANDiscovery,Jacob2025SPIKAN}. Whether their edge-based structure supports tomography-specific inspection remains an open question. This edge-based structure allows channel selection, pruning-compatible pathway analysis, hidden-representation probes, and sparse formula recovery to be combined within a single model-inspection workflow.

This architectural distinction motivates our use of a KAN rather than an MLP as the primary inspectable model. After pruning, a KAN provides named input-hidden-output pathways that can be ranked directly in Pauli-channel units. An MLP can still be interrogated by external ablation and hidden-state probes, but its internal routes are not native model objects and require post-hoc attribution proxies. We therefore use the MLP as an architecture-level reference, while the sparsified KAN supplies the pathway object needed for tomography-specific inspection.

To evaluate this sparsified-KAN inspection framework, we choose the three-qubit GHZ family as a controlled testbed because its states lie in the subspace spanned by $\ket{000}$ and $\ket{111}$. Moreover, its population imbalance and off-diagonal components have closed-form expressions in a small set of Pauli expectation values~\cite{Greenberger1990,Mermin1990,Pan2000}. This task is not a general-purpose three-qubit tomography problem and does not reconstruct a complete $8 \times 8$ density matrix. Instead, it asks whether a sparsified KAN trained on all 63 non-identity Pauli expectation values can recover a physically meaningful GHZ-subspace reconstruction rule.

From a machine-learning methodology perspective, the study focuses not only on reconstruction accuracy but also on whether the learned reconstruction rule can be inspected against known physical structure. We use the analytically tractable GHZ-subspace tomography problem as a controlled setting for testing whether a neural reconstruction model can be inspected against known physical structure. Our contributions are fourfold:

\begin{enumerate}
    \item We introduce sparsified KANs as an inspection-oriented AI/ML framework for quantum state tomography. This opens a direction beyond conventional fidelity optimization: the learned reconstruction rule can be audited through sparse edge functions, retained Pauli channels, and pathway-level structure.

    \item We bridge Pauli-resolved quantum-state reconstruction and Kolmogorov--Arnold Network model inspection. The GHZ-subspace setting links physical observables, namely Pauli expectation values, with machine-learning objects, namely edge functions, hidden units, pruning patterns, and output-level formulas.

    \item We deliver a reproducible inspection toolkit for neural tomography. The workflow combines Pauli-channel ablation, pruning-compatible pathway analysis, hidden-subspace probes, sparse formula recovery, sparse linear baselines, selected-channel retraining, and negative controls.

    \item We provide a mechanistic account of the trained model's behavior. The analyses show which Pauli measurements the KAN uses, how hidden units mediate physically meaningful channel groups, how pruning recovers measurement combinations consistent with analytic GHZ relations, and where this GHZ-subspace rule breaks down under off-manifold stress tests.
\end{enumerate}

The remainder of the paper is organized as follows. Section~II first defines the sparsified KAN tomography framework and then introduces the GHZ-subspace benchmark, data-generation protocol, post-training inspection analyses, baselines, controls, and evaluation metrics. Section~III first establishes reconstruction accuracy under controlled noise, then tests whether the learned rule recovers the expected Pauli channels, internal routing, hidden representations, and signed formulas. It also uses sparse linear baselines and control experiments to define the scope of the interpretation. Section~IV discusses the resulting evidence chain, its limitations, and possible extensions.

\section{Methods}

\subsection{Interpretable KAN tomography framework}

We use a sparsified KAN as an inspectable neural reconstruction model rather than as an opaque regressor. The model maps estimated Pauli expectation values to reconstructed target variables, while its retained edges, hidden units, and output-level summaries provide objects for post-training inspection. We formulate this reconstruction map as
\begin{equation}
    \hat{\mathbf r}=f_{\mathrm{KAN}}(\hat{\mathbf m}),\qquad
    \hat{\mathbf m}\in\mathbb{R}^{63},\qquad
    \hat{\mathbf r}=(\hat z,\hat c,\hat s).
    \label{eq:kan-tomography-map}
\end{equation}
Here $\mathbf m$ denotes the ideal Pauli-expectation vector, $\hat{\mathbf m}$ denotes the corresponding finite-shot or noise-affected estimate, and $\hat{\mathbf r}$ denotes the KAN prediction. The trained model is evaluated both for reconstruction accuracy and as an inspectable reconstruction rule. Its input dependence, internal organization, hidden representation, and output-level formula summaries are compared with the known Pauli structure of the benchmark state family.

\begin{figure*}[t]
\centering
\includegraphics[width=\textwidth]{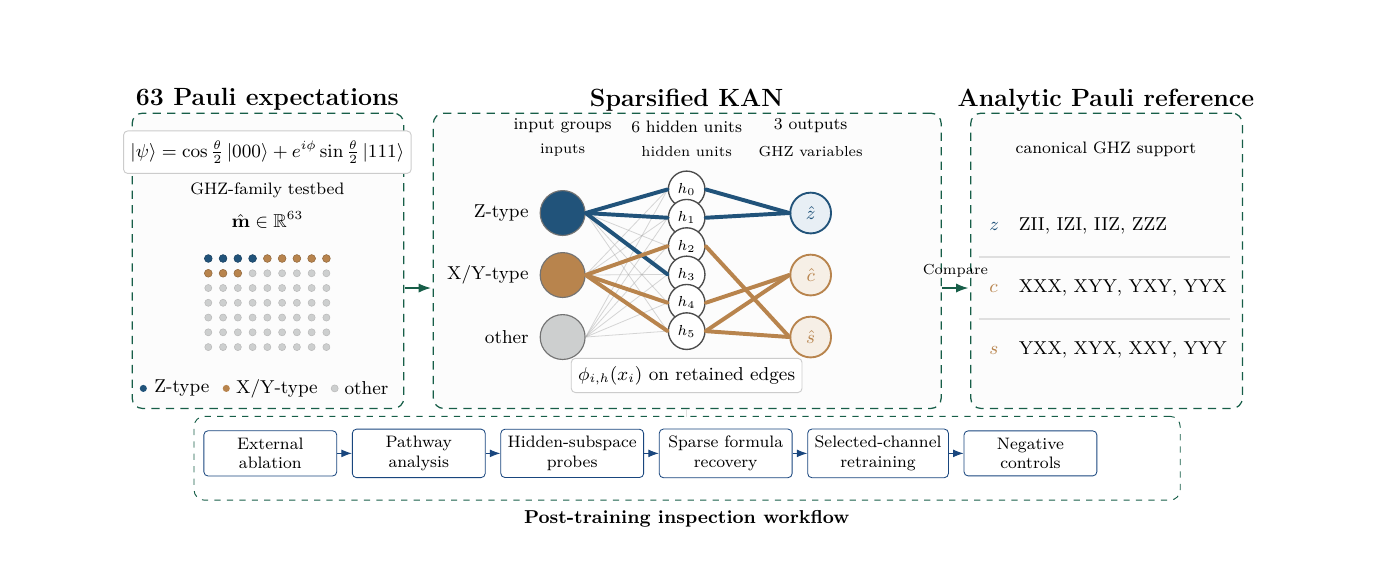}
\caption{\label{fig:kan-framework}
Sparsified KAN tomography workflow. A full 63-channel Pauli-expectation vector is mapped to the GHZ-subspace variables $(z,c,s)$ by a compact KAN and compared with the analytic GHZ Pauli reference. After training and pruning, the retained model is inspected through external ablation, pathway analysis, hidden-subspace probes, sparse Pauli formula recovery, selected-channel retraining, and negative controls.}
\end{figure*}

Figure~\ref{fig:kan-framework} summarizes the resulting inspection workflow, from Pauli-expectation inputs and sparse KAN reconstruction to post-training channel, pathway, hidden-subspace, and formula-level probes.

The primary neural model is a KAN with 63 input channels, six hidden units, and three output variables. In a KAN, learnable one-dimensional functions are assigned to edges rather than to nodes~\cite{Liu2024KAN,Liu2024KAN2}. After training and pruning, as described below, this edge-based representation yields a sparse computational graph whose pathways link named Pauli observables to reconstructed variables through individual hidden units. We use these pathways to test whether the learned rule separates Z-type population observables from X/Y off-diagonal observables and routes them to the corresponding outputs $z$, $c$, and $s$. The purpose of using a KAN in this study is therefore not to claim superior sparse regression on this benchmark, but to obtain pathway-level structural interpretability within a neural tomography model.

\subsection{GHZ-subspace benchmark and Pauli-expectation representation}

We use the three-qubit GHZ family~\cite{Greenberger1990} as a controlled analytic testbed for auditing the inspection workflow. The family is useful here because its population imbalance and off-diagonal coherences have known sparse Pauli representations. This design allows the learned reconstruction rule to be compared with an analytic reference structure. It is not intended as a general-purpose benchmark for arbitrary three-qubit quantum state tomography. The pure GHZ-family states are parameterized as
\begin{equation}
    \ket{\psi(\theta,\phi)}
    =
    \cos\frac{\theta}{2}\ket{000}
    +
    e^{i\phi}\sin\frac{\theta}{2}\ket{111},
    \label{eq:ghz-family}
\end{equation}
where $\theta\in[0,\pi]$ and $\phi\in[-\pi,\pi]$. The reconstruction target is the vector $\mathbf r=(z,c,s)$, whose components are defined from the computational-basis amplitudes by
\begin{align}
    z &= |\psi_{000}|^2 - |\psi_{111}|^2,\\
    c &= 2\,\mathrm{Re}(\psi_{000}\psi_{111}^{*}),\\
    s &= 2\,\mathrm{Im}(\psi_{000}\psi_{111}^{*}).
\end{align}
For the state in Eq.~\eqref{eq:ghz-family}, these definitions give
\begin{equation}
    z=\cos\theta,\qquad
    c=\sin\theta\cos\phi,\qquad
    s=-\sin\theta\sin\phi.
\end{equation}
The sign convention for $s$ is fixed by writing the off-diagonal product as $\psi_{000}\psi_{111}^{*}=(c+is)/2$. The density-matrix elements within this subspace are then
\begin{align}
    \rho_{000,000} &= \frac{1+z}{2},&
    \rho_{111,111} &= \frac{1-z}{2},\\
    \rho_{000,111} &= \frac{c+is}{2},&
    \rho_{111,000} &= \frac{c-is}{2}.
\end{align}

Each input vector contains all 63 non-identity three-qubit Pauli expectation values, following the standard Pauli-expectation representation used in qubit tomography~\cite{James2001,ParisRehacek2004,Qi2013}. Unless stated otherwise, models are trained on the full 63-channel input rather than on a preselected GHZ subset. For post-training evaluation, we define the following extended GHZ-relevant Pauli set as the reference support for overlap metrics and sparse formula recovery. The same set is also used as a restricted input subset in selected-channel retraining:
\begin{equation}
\begin{split}
\mathcal{P}_{\mathrm{GHZ}}^{(12)}=\{&
\mathrm{ZII}, \mathrm{IZI}, \mathrm{IIZ}, \mathrm{ZZZ}, \mathrm{XXX}, \mathrm{XYY},\\
&\mathrm{YXY}, \mathrm{YYX}, \mathrm{YXX}, \mathrm{XYX}, \mathrm{XXY}, \mathrm{YYY}\}.
\end{split}
\end{equation}
Throughout, each Pauli string denotes the corresponding expectation value. The analytic GHZ Pauli relations are
\begin{align}
    z &= \frac{1}{4}(\mathrm{ZII} + \mathrm{IZI} + \mathrm{IIZ} + \mathrm{ZZZ}), \label{eq:z-pauli}\\
    c &= \frac{1}{4}(\mathrm{XXX} - \mathrm{XYY} - \mathrm{YXY} - \mathrm{YYX}), \label{eq:c-pauli}\\
    s &= \frac{1}{4}(-\mathrm{YXX} - \mathrm{XYX} - \mathrm{XXY} + \mathrm{YYY}). \label{eq:s-pauli}
\end{align}
These formulas are not supplied as training constraints. Instead, they provide post-training reference structure for input-channel validation and sparse formula recovery. We also use a smaller primary GHZ set,
\begin{equation}
    \mathcal{P}_{\mathrm{GHZ}}^{(7)}
    =
    \{\mathrm{ZII}, \mathrm{IZI}, \mathrm{IIZ}, \mathrm{XXX}, \mathrm{YXX}, \mathrm{XYX}, \mathrm{YYX}\},
\end{equation}
in selected-channel retraining to assess redundancy in the extended representation.
The complete 63-channel Pauli ordering is provided in Appendix~\ref{app:ghz-benchmark}.

\subsection{Data generation, noise models, and model training}

\textbf{Data generation.}
For the GHZ-family data set, $\theta$ and $\phi$ are sampled independently from uniform distributions on $[0,\pi]$ and $[-\pi,\pi]$, respectively. The default data split contains 1400 training samples, 300 validation samples, and 300 test samples. Unless stated otherwise, noise sweeps and stability analyses are repeated over seeds 0--4. Extended supplementary controls use seeds 0--9. The hidden-subspace alignment analysis uses 20 random initializations (seeds 0--19) at the 1000-shot condition.

\textbf{Noise models.}
Finite-shot measurement noise is simulated independently for each Pauli observable. For an ideal Pauli expectation value $m$, each measurement shot returns a $\pm 1$ outcome with probabilities $P(+1)=(1+m)/2$ and $P(-1)=(1-m)/2$. The noisy input expectation value is then estimated from shot counts of 500, 1000, 5000, or 10000. In this finite-shot protocol, the Pauli-expectation inputs are noisy estimates, whereas the targets remain the clean coordinates $(z,c,s)$ of the underlying state.

Depolarizing noise is represented by scaling all non-identity Pauli expectations and the corresponding target variables by $1-p$, with $p = 0.01, 0.05, 0.10, 0.20$. Because the targets are also attenuated in this protocol, a model can achieve high fidelity to the noisy targets without reconstructing the original pure GHZ-family state. We therefore report two fidelity measures for depolarizing experiments: noisy-target fidelity, evaluated against the attenuated target variables, and clean-target fidelity, evaluated against the original pure-GHZ variables.

\textbf{Model training.}
All neural models are trained to minimize mean-squared error on the three targets $(z,c,s)$. The KAN uses grid size 5 and spline order 3. The training pipeline consists of Adam optimization, pruning, and LBFGS fine-tuning of the pruned model, followed by test-set evaluation and post-training inspection. Full optimizer settings, regularization settings, and stopping details are provided in Appendix~\ref{app:training-protocols}. Pruning removes nodes and edges whose pykan importance scores fall below a specified threshold; the default threshold, safe-pruning procedure, and input-index mapping checks are described in Appendix~\ref{app:training-protocols}. The pruning-threshold sensitivity analysis is reported in Appendix~\ref{app:control-diagnostics}. The safe-pruning routine preserves at least one hidden node in cases where small selected-channel runs would otherwise remove the intermediate layer.

\subsection{Post-training KAN inspection analyses}

The post-training analyses audit complementary aspects of the learned reconstruction rule. External ablation measures input dependence. Pathway analysis measures internal organization and routing magnitude in the sparse KAN graph. Hidden-subspace probes test hidden-representation alignment with the target variables. Sparse formula recovery tests signed formula consistency with the analytic Pauli relations.

\textbf{External channel ablation.} External channel-ablation importance is computed by zeroing one Pauli input channel at a time and measuring the resulting increase in target-variable RMSE. This perturbation-based score provides an input-level test of whether the trained model relies on the Pauli observables that analytically encode $z$, $c$, and $s$. It is used as an external dependence measure rather than as a direct measure of internal routing.

\textbf{Pathway analysis.} Post-pruning input indices are mapped back to the original Pauli labels so that each pathway linking a Pauli observable to an output variable through a hidden unit is assigned to a physical observable. Pathway scores are computed as the product of the absolute input--hidden and hidden--output KAN edge strengths. For each output variable, all pathways are ranked by this score, and the $k$ highest-scoring pathways are retained. The top-$k$ target-support fraction is the fraction of these rows whose input Pauli channel belongs to the canonical support of that output. This metric is computed over pathway rows rather than over unique input channels, and it measures routing magnitude rather than algebraic sign. The target-specific supports are $\{\mathrm{ZII},\mathrm{IZI},\mathrm{IIZ},\mathrm{ZZZ}\}$ for $z$, $\{\mathrm{XXX},\mathrm{XYY},\mathrm{YXY},\mathrm{YYX}\}$ for $c$, and $\{\mathrm{YXX},\mathrm{XYX},\mathrm{XXY},\mathrm{YYY}\}$ for $s$. Aggregated pathway scores are obtained by summing over hidden units,
\begin{equation}
    \mathrm{score}_{i,o}=\sum_h |\mathrm{scale}_{i,h}|\,|\mathrm{scale}_{h,o}|.
\end{equation}
Pathway scores quantify routing importance but do not encode algebraic sign; signed Pauli structure is assessed by sparse formula recovery.

\textbf{Hidden-subspace probes.} Hidden-subspace probes test whether the learned internal representation contains the target variables in a linearly accessible form. Hidden-unit correlations are computed as the absolute Pearson correlation between each hidden unit's activation and each target variable $(z,c,s)$ on held-out test samples. These single-unit diagnostics do not require a one-to-one assignment between hidden units and physical variables. To test subspace-level alignment, canonical correlation analysis (CCA) is applied to the $N \times 6$ matrix of hidden activations and the $N \times 3$ matrix of target variables on held-out test samples; canonical correlations are reported for each component. Linear probes are trained on training-set hidden activations and evaluated on held-out test activations.

\textbf{Sparse formula recovery.} Sparse formula recovery tests whether the KAN outputs admit signed Pauli summaries consistent with the analytic relations. Candidate support is selected from all 63 Pauli channels using the absolute correlation between each Pauli input channel and the corresponding KAN-predicted output variable. The selection uses a relative threshold of 0.95 times the maximum correlation and is capped at 12 terms. Least-squares fitting is then performed on the selected support, followed by coefficient thresholding at $10^{-3}$~\cite{Schmidt2009,Brunton2016,Rudy2017,Cranmer2020}. Formula recovery from the true target variables uses the same protocol for baseline comparisons. Because the highly symmetric GHZ manifold contains redundant Pauli information, the recovered support should be interpreted as convention-dependent rather than unique.

\subsection{Baselines, controls, and evaluation metrics}

\textbf{Linear baselines.}
The target variables have sparse linear representations in the Pauli basis (Eqs.~\eqref{eq:z-pauli}--\eqref{eq:s-pauli}). Linear models therefore provide formula-level baselines for separating predictive accuracy and sparse support recovery from KAN-specific pathway-level neural interpretability. We evaluate ordinary least squares (OLS), ridge regression~\cite{HoerlKennard1970}, LASSO~\cite{Tibshirani1996}, and elastic net~\cite{ZouHastie2005} using the same Pauli-expectation inputs, data splits, and default 1000-shot finite-shot condition as the KAN. The same models are also evaluated under noiseless and purely depolarized settings. OLS and ridge are used as dense linear prediction references, whereas LASSO and elastic net are used as sparse support-recovery references. LASSO and elastic-net regularization strengths are selected independently for each output variable on the validation set; the elastic-net mixing parameter is fixed at 0.8. Coefficients are standardized during fitting and transformed back to Pauli-channel units before support and coefficient comparisons. For each linear model, the 12 largest-magnitude coefficients are identified by pooling across the three output-specific coefficient vectors and ranking by absolute magnitude. A coefficient threshold of $10^{-3}$ defines recovered support. After hyperparameter selection, linear models are refit on the combined training and validation sets before test evaluation.

\textbf{Neural baseline.}
The neural reference model is a width-matched bottleneck MLP with 63 input channels, six hidden units, and three output variables, using ReLU activation, Adam training for 300 epochs, and learning rate 0.01. The MLP is evaluated with the same external channel-ablation and hidden-subspace probe analyses as the KAN. This comparison is used as an architecture-level reference for structural readability rather than as an exhaustive benchmark of neural tomography models. Full hyperparameters, training details, and the post-hoc input-hidden attribution proxy used for the MLP internal support score are provided in Appendix~\ref{app:kan-mlp-comparison}.

\textbf{Control analyses.}
Control analyses test whether the interpretation is robust to feature-order changes, irrelevant channels, random labels, restricted-input retraining, and off-manifold perturbations. The shuffled input-order check retrains the model after randomly permuting Pauli input columns and then maps the resulting rankings back to Pauli names. The decoy-channel control appends Gaussian, shuffled-Pauli, and variance-matched channels, each matched in variance to the corresponding Pauli channel in the training set, and tests whether any decoy enters the top-20 ranked set. The random-label negative control shuffles the training input--target pairing while evaluation is performed on unshuffled held-out data. Noise-level Jaccard support stability measures the overlap of recovered Pauli support across finite-shot and depolarizing noise levels.

Additional controls test functional sufficiency and off-manifold boundary behavior. For selected-channel retraining, KANs are retrained from scratch with the same training protocol. The KAN-selected top-12 subset is obtained from the external channel-ablation ranking of the all-63 model. Random 12-channel subsets serve as negative controls, with 30 total runs from ten random draws per seed across seeds 0--2. In the 90/10 training experiment, the training set contains 1400 samples, consisting of 1260 GHZ-family samples and 140 Haar-random three-qubit pure states. The validation set contains 300 GHZ-family samples, and the test set contains 400 GHZ-family samples. This experiment is an independent three-seed control rather than a rerun of the main noise sweep. The 90/10 and W-class stress-test controls both use 400 test samples. For off-manifold states, the target variables are obtained from the same $\ket{000}$/$\ket{111}$ density-matrix block used to define $(z,c,s)$ above, rather than from a full-state target. The W-class stress test evaluates GHZ-trained models on controlled mixtures of GHZ-family and W-class states. The W-class component is supported on $\ket{001}$, $\ket{010}$, and $\ket{100}$, with relative phases drawn uniformly from $[0,2\pi)$. The mixtures are constructed at the density-matrix level, with tested W-class mixture fractions of $0$, $0.05$, and $0.10$. Additional control analyses evaluate random-initialization stability and sparsity-threshold sensitivity.

\textbf{Evaluation metrics.}
Reconstruction error is reported as root-mean-squared error (RMSE) on the target vector $\mathbf r=(z,c,s)$. Component-wise RMSE is computed for $z$, $c$, and $s$ over held-out samples; when a single RMSE value is reported, it denotes the mean of these three. Before computing reconstruction metrics, both predicted and target vectors are projected to the physical Bloch ball by clipping the norm to 1 whenever $\|\hat{\mathbf r}\| > 1$ or $\|\mathbf r\| > 1$. This projection is applied uniformly to predictions and targets so that fidelity and RMSE are evaluated in the same physical representation. For two vectors $\mathbf r=(z,c,s)$ and $\hat{\mathbf r}=(\hat z,\hat c,\hat s)$, the fidelity is
\begin{equation}
F(\mathbf r,\hat{\mathbf r})=\frac{1}{2}\left[
1+\mathbf r\cdot\hat{\mathbf r}+
\sqrt{(1-\|\mathbf r\|^2)(1-\|\hat{\mathbf r}\|^2)}
\right].
\end{equation}
Unless stated otherwise, all reported fidelities are computed within this subspace. They are not full $8\times8$ density-matrix fidelities for arbitrary three-qubit states.

Support overlap is quantified using the Jaccard index,
\begin{equation}
J(A,B)=\frac{|A\cap B|}{|A\cup B|},
\end{equation}
between a recovered Pauli-support set $A$ and the corresponding GHZ reference support $B$.

The main channel-support overlap metrics are defined as follows. The top-12 extended-channel Jaccard compares the 12 highest-ranked channels, ranked by external ablation or coefficient magnitude, with $\mathcal{P}_{\mathrm{GHZ}}^{(12)}$. The retained-set Jaccard compares the post-pruning retained input set with the same reference set. Target-level support precision and recall compare recovered formula terms with the target-specific canonical supports in Eqs.~\eqref{eq:z-pauli}--\eqref{eq:s-pauli}.

For hidden linear probes, the coefficient of determination is computed on held-out samples as
\begin{equation}
R^2 = 1 - \frac{\sum_i \|y_i - \hat{y}_i\|^2}{\sum_i \|y_i - \bar{y}\|^2},
\end{equation}
with sums taken over both samples and output components. For sparse formula recovery, target-wise $R^2$ is computed with the same variance-explained definition on the formula-recovery evaluation samples.

\section{Results}

\subsection{Reconstruction accuracy and noise stability}

\begin{figure*}[!t]
\centering
\includegraphics[width=\textwidth]{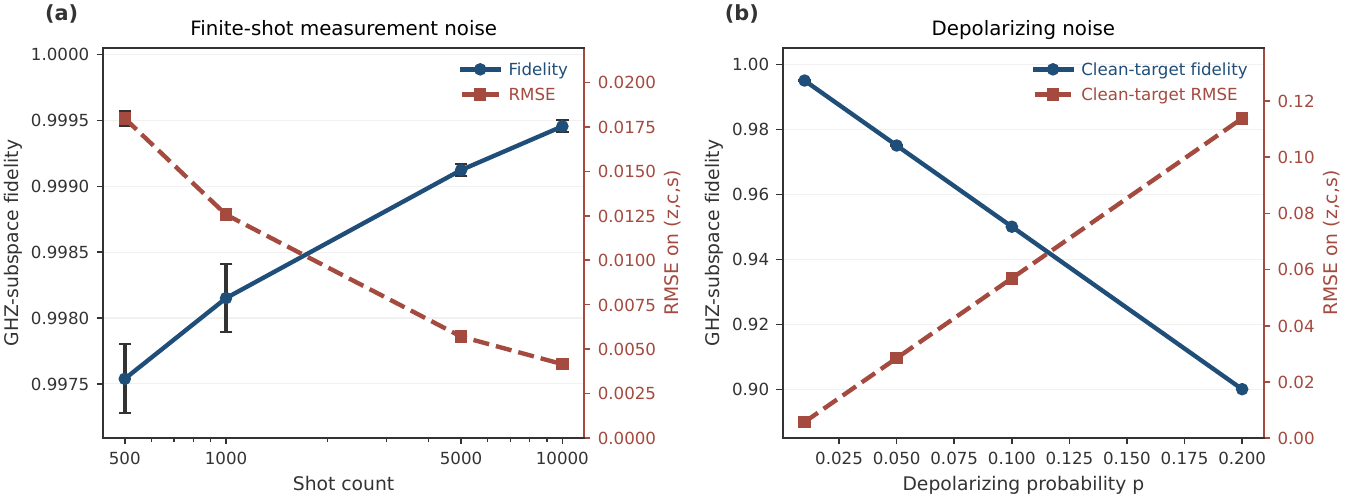}
\caption{\label{fig:reconstruction-noise}
Reconstruction of GHZ-subspace variables under finite-shot and depolarizing noise. (a) Finite-shot measurement noise: GHZ-subspace fidelity (left axis) and RMSE over $\ghzvars$ (right axis) versus shot count. (b) Depolarizing noise: clean-target fidelity (left axis) and clean-target RMSE (right axis) versus depolarizing probability. Line samples and curve labels are shown in the upper-right corner of each panel. All values are averaged over seeds 0--4; error bars denote $\pm$ one standard deviation.}
\end{figure*}

Before analyzing what the KAN learns, we first ask whether the reconstruction accuracy is sufficient for model inspection to be meaningful. Figure~\ref{fig:reconstruction-noise} therefore establishes an accuracy baseline rather than a performance-comparison claim.

Under finite-shot noise, the GHZ-subspace fidelity increased from $0.997538 \pm 0.000263$ at 500 shots to $0.999455 \pm 0.000045$ at 10000 shots. Over the same range, the mean RMSE over $\ghzvars$ decreased from $0.017974 \pm 0.000408$ to $0.004155 \pm 0.000080$. At the default 1000-shot setting, the clean-target fidelity was $0.998151 \pm 0.000257$, with mean RMSE $0.012562 \pm 0.000322$. The improvement with increasing shot count is consistent with the $1/\sqrt{N}$ scaling of statistical uncertainty in the Pauli expectation estimates.

The depolarizing-noise sweep distinguishes two fidelity notions. For depolarizing noise, both Pauli inputs and target labels are scaled by $1-p$. The resulting noisy-target fidelity remained near unity across the tested range, from $0.999998 \pm 0.000001$ at $p=0.01$ to $1.000000 \pm 0.000000$ at $p=0.20$. By contrast, clean-target fidelity relative to the original pure GHZ target decreased from $0.994998 \pm 0.000009$ at $p=0.01$ to $0.900034 \pm 0.000015$ at $p=0.20$, and clean-target RMSE increased from $0.005720 \pm 0.000030$ to $0.113814 \pm 0.000434$; these clean-target trends are shown in Fig.~\ref{fig:reconstruction-noise}b.

Thus, Fig.~\ref{fig:reconstruction-noise} establishes that the reconstruction task is solved accurately enough for subsequent inspection analyses to be meaningful. The following sections therefore focus on which Pauli channels the model uses, how those channels are organized internally, and whether the resulting interpretation survives controls (Secs.~III.B--III.E).

\subsection{Input-level Pauli channel identification}

Having established the accuracy baseline, we next ask whether the trained KAN identifies the Pauli observables that physically encode the GHZ-subspace variables, and whether this identification is robust to column-order artifacts and statistically plausible but irrelevant input channels.

\begin{figure*}[!t]
\centering
\includegraphics[width=\textwidth]{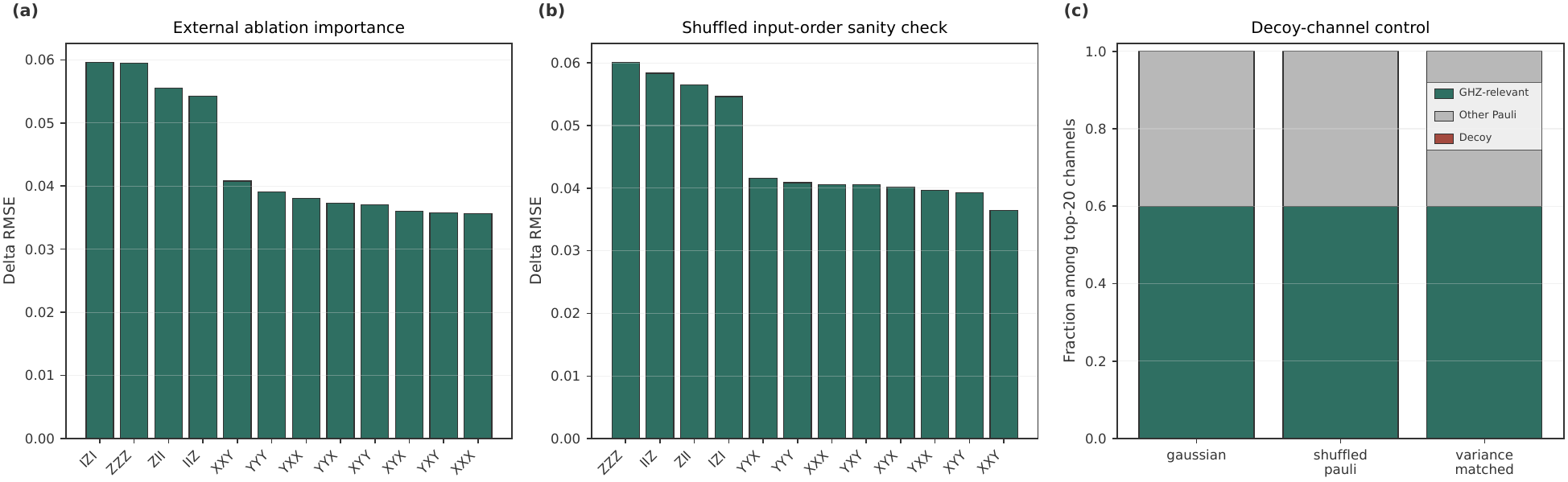}
\caption{\label{fig:input-interpretability}
Input-level analyses identify GHZ-relevant Pauli channels. All panels use the 1000-shot condition. (a) Mean external-ablation importance, defined as the increase in clean-target RMSE after zeroing each Pauli channel and averaged over seeds 0--4; canonical GHZ-relevant measurements appear as top-ranked channels. (b) Mean external-ablation importance after random permutation of the 63 input columns and retraining, averaged over shuffled seeds 0--2 and mapped back to the original Pauli labels; the same GHZ-relevant channels remain top-ranked. (c) Decoy-channel control: after appending 24 Gaussian, shuffled-Pauli, or variance-matched decoy channels and recomputing top-20 external-ablation rankings (nine seed-by-decoy runs total), stacked bars show the fractions of GHZ-relevant true channels, other true Pauli channels, and decoy channels among the top-20 ranked channels. The decoy component is not visible because no appended decoy channel entered the top-20 ranking in any of the nine seed-by-decoy control runs, yielding a mean decoy fraction of $0.000 \pm 0.000$.}
\end{figure*}

Figure~\ref{fig:input-interpretability}a shows the mean 1000-shot external-ablation importance ranking, averaged over seeds 0--4. For a representative seed-0 run at 1000 shots, external ablation recovered exactly the extended GHZ-relevant top-12 set, with top-12 extended-channel Jaccard 1.000. The four strongest channels were the Z-type population observables ($\mathrm{ZZZ}$, $\mathrm{IZI}$, $\mathrm{ZII}$, $\mathrm{IIZ}$), followed by the eight X/Y off-diagonal observables. The absolute $\Delta$RMSE values are small because the baseline reconstruction error is already low: at 1000 shots, the mean RMSE is $0.012562$. Thus, an ablation-induced increase of order $10^{-2}$ represents a change comparable to the baseline error itself. These rankings indicate that the KAN depends primarily on the 12 GHZ-relevant Pauli channels, with Z-type population observables dominating the importance scores.

The selected channels were not determined by column position. Figure~\ref{fig:input-interpretability}b shows the mean 1000-shot external-ablation importance after randomly permuting the 63 Pauli input columns, retraining across shuffled seeds 0--2, and mapping the resulting rankings back to the original Pauli labels. Across these three shuffled-input retraining runs, the top-12 extended-channel Jaccard remained $1.000 \pm 0.000$. All 12 top-ranked channels were GHZ-relevant in every seed, and the mean clean-target fidelity was $0.998199 \pm 0.000036$.

The decoy-channel control tests whether irrelevant but statistically plausible inputs enter the top-ranked set. Figure~\ref{fig:input-interpretability}c summarizes the composition of the top-20 ranked channels after appending 24 decoy channels of three types: Gaussian, shuffled-Pauli, and variance-matched. Each stacked bar reports the fractions of GHZ-relevant true channels, other true Pauli channels, and decoy channels among the top-20 external-ablation ranking, averaged over seeds for that decoy type. In all three decoy settings, the mean decoy fraction was $0.000 \pm 0.000$. Across all nine seed-by-decoy runs, the top 20 contained 12 GHZ-relevant true channels, 8 other true Pauli channels, and no decoy channels. The mean clean-target fidelity in this augmented-input control was $0.998064 \pm 0.000290$. Together, the shuffled-input and decoy-channel controls make column-order and irrelevant-channel artifacts unlikely explanations for the KAN's channel preference. The result therefore supports interpreting Fig.~\ref{fig:input-interpretability} as Pauli-channel identification from the full 63-channel input, not merely as a feature-importance ranking.

\subsection{Internal organization: pathways, hidden subspace, and formula recovery}

The input-level analysis identifies which Pauli channels the KAN depends on. We next ask whether these channels are organized inside the sparse KAN in a way that is consistent with the GHZ Pauli structure: Z-type population observables associated with $z$, and X/Y off-diagonal observables associated with $c$ and $s$.

\textbf{Pathway routing.}
Pathway analysis asks whether the Pauli support identified at the input level is organized inside the sparse KAN in an output-specific way. In the representative seed-0 1000-shot model, the highest-scoring pathways for the population output $z$ are drawn from the Z-type population support, whereas the two off-diagonal outputs concentrate their strongest pathways on the corresponding X/Y Pauli support groups. The top-4 target-support fraction is 1.000 for all three outputs; at larger $k$, lower-ranked pathways include shared or cross-output channels, so the pathway analysis is interpreted as evidence for dominant-route organization rather than perfect separation of all retained routes.

\begin{figure*}[!t]
\centering
\includegraphics[width=\textwidth]{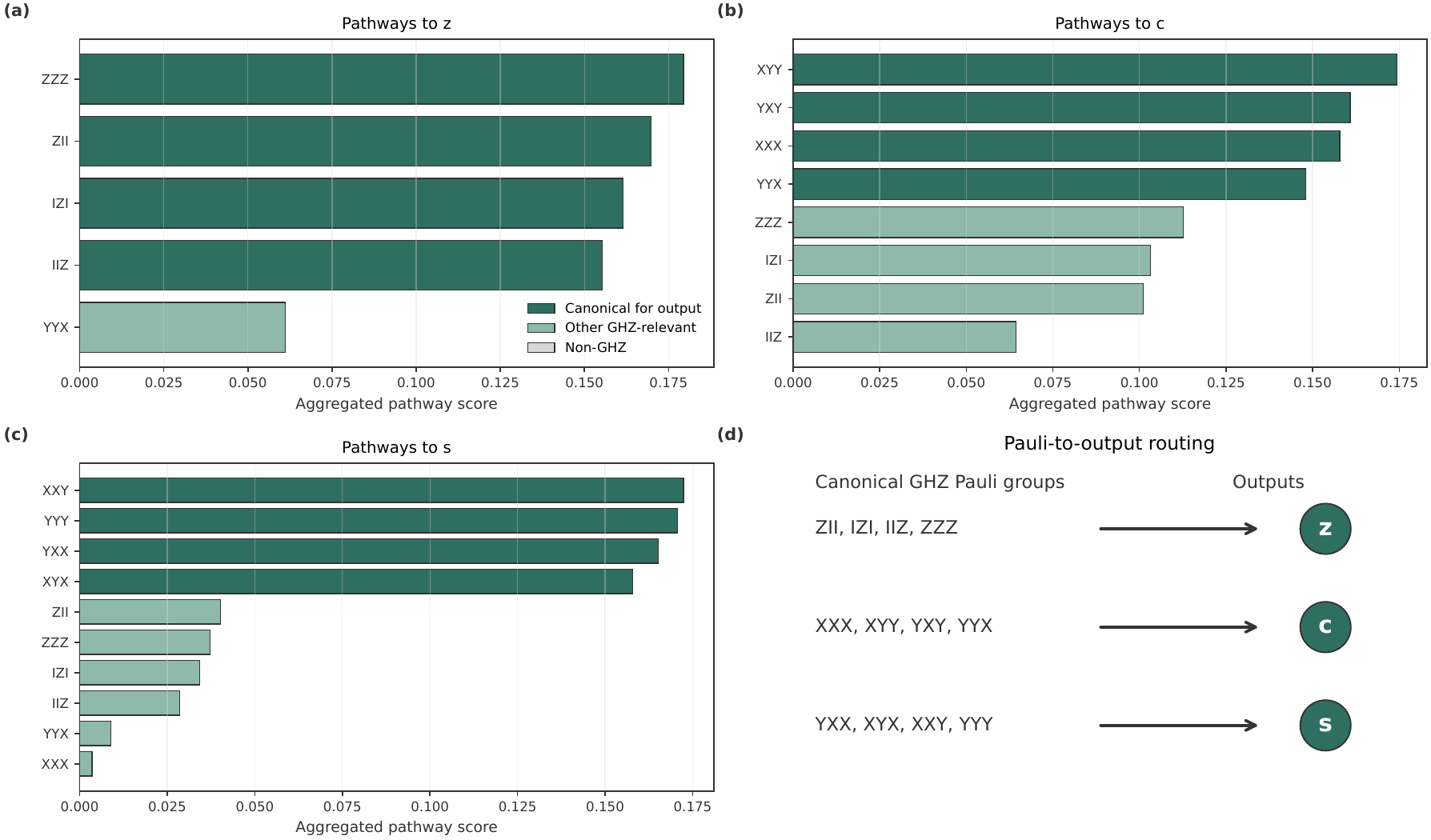}
\caption{\label{fig:mechanism-interpretability}
Aggregated pathway scores for each output variable, obtained by summing $|\mathrm{scale}_{i,h}|\cdot|\mathrm{scale}_{h,o}|$ over hidden units. (a--c) Top Pauli channels ranked by aggregated score to $z$, $c$, and $s$, respectively. Dark green bars denote canonical GHZ-relevant channels for the displayed output, pale green denotes other GHZ-relevant channels, and gray denotes non-GHZ channels. (d) Routing schematic for the canonical Pauli groups associated with the population, real off-diagonal, and imaginary off-diagonal outputs. Scores use absolute edge strengths and quantify routing importance without encoding algebraic sign. All panels are from a representative seed-0 KAN at the 1000-shot condition.}
\end{figure*}

Figure~\ref{fig:mechanism-interpretability} summarizes this organization using aggregated pathway scores, $\mathrm{score}_{i,o}=\sum_h |\mathrm{scale}_{i,h}|\,|\mathrm{scale}_{h,o}|$. The aggregated scores concentrate each output on the canonical Pauli group associated with that GHZ variable. Because these scores use absolute edge strengths, they quantify routing magnitude rather than algebraic sign; signed structure is tested separately by sparse formula recovery. These patterns are consistent with pathway-level internal organization, without by themselves establishing a causal mechanism.

\begin{figure*}[!t]
\centering
\includegraphics[width=\textwidth]{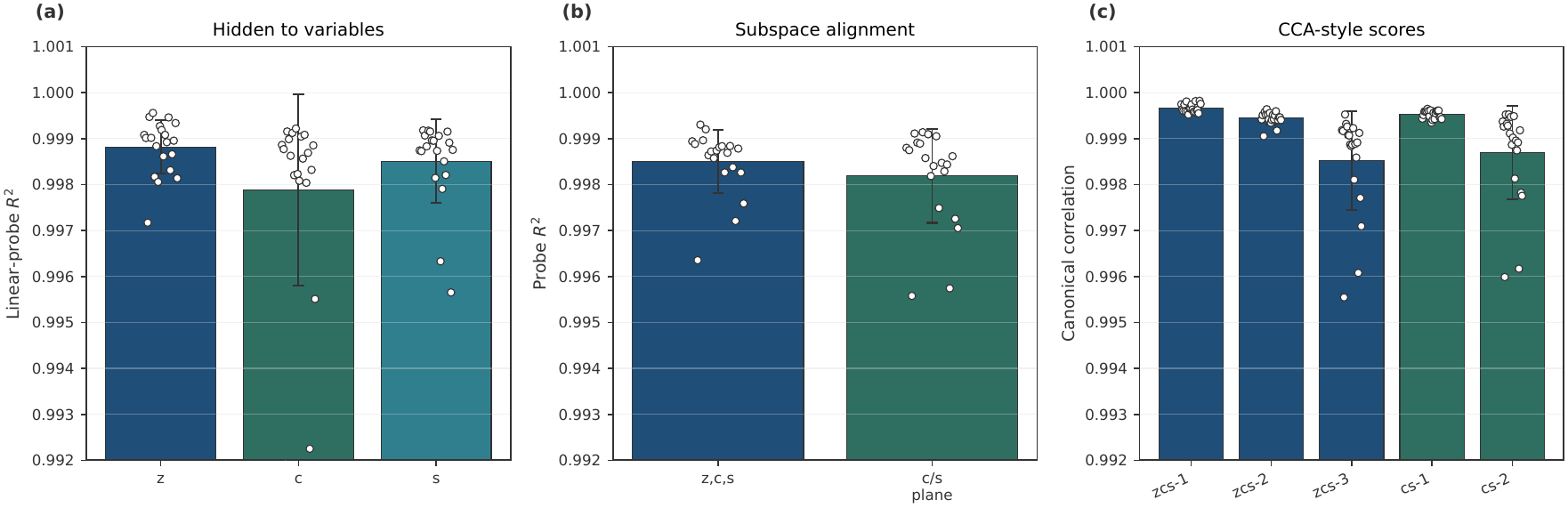}
\caption{\label{fig:hidden-subspace-interpretability}
Hidden-subspace alignment metrics across 20 random initializations at 1000 shots. Bars show held-out $R^2$ from linear probes trained on hidden activations (targets $z$, $c$, $s$, the $(c,s)$ plane, and the full $(z,c,s)$ vector), together with canonical correlations from CCA-style alignment (first component of the full space, and the two components of the $(c,s)$ plane). Error bars denote $\pm$ one standard deviation across seeds. The vertical axes are zoomed to show small variations.}
\end{figure*}

\textbf{Hidden-subspace probes.} Single-neuron correlations provide an initial diagnostic of hidden-variable alignment: for each GHZ variable $v \in \{z,c,s\}$, we compute $\max_h |\mathrm{corr}(h,v)|$ over hidden units. Across 20 random initializations at 1000 shots, the mean maximum absolute correlations were $0.971330 \pm 0.051790$ for $z$, $0.857943 \pm 0.155109$ for $c$, and $0.845563 \pm 0.164782$ for $s$, with the off-diagonal variables showing larger seed-to-seed variability in their maximum single-neuron correlations than $z$. Linear probes test whether the GHZ variables can be decoded from a weighted combination of all hidden activations; a least-squares linear readout trained on training-set hidden activations and evaluated on held-out test activations achieved $R^2 = 0.9985 \pm 0.0007$ for the full $(z,c,s)$ vector and $R^2 = 0.9982 \pm 0.0010$ for the $(c,s)$ plane across 20 initializations. CCA-style alignment measures subspace overlap between hidden activations and GHZ variables without requiring a preassigned hidden-unit-to-variable correspondence; canonical correlations exceeded 0.99 for both the full $(z,c,s)$ space and the $(c,s)$ plane (Table~\ref{tab:S2}). Figure~\ref{fig:hidden-subspace-interpretability} summarizes these hidden-subspace probes, and Table~\ref{tab:S2} reports the full component-wise and CCA statistics. Taken together, these hidden-representation probes indicate that the KAN's internal activations contain a linearly accessible representation of the GHZ-subspace variables, without requiring a one-to-one assignment between individual hidden units and physical variables.

\textbf{Sparse formula recovery.} Sparse formula recovery tests whether the KAN outputs can be summarized by the signed Pauli relations, using support selected from all 63 Pauli channels rather than a pre-specified template fit~\cite{Schmidt2009,Brunton2016,Rudy2017,Cranmer2020}. Applied to KAN outputs, this procedure recovered the canonical GHZ supports with support precision and recall both equal to 1.000 for all three variables.

The recovered KAN-output formulas were
\begin{equation}
\begin{aligned}
    z &\approx 0.2500\,
    (\mathrm{ZII}+\mathrm{IZI}+\mathrm{IIZ}+\mathrm{ZZZ}),\\
    c &\approx 0.2500\,
    (\mathrm{XXX}-\mathrm{XYY}-\mathrm{YXY}-\mathrm{YYX}),\\
    s &\approx 0.2500\,
    (-\mathrm{YXX}-\mathrm{XYX}-\mathrm{XXY}+\mathrm{YYY}).
\end{aligned}
\end{equation}
For $z$, $c$, and $s$, respectively, the KAN-output sparse fits achieved $R^2 = 0.999999768$, 0.999999602, and 0.999999630, with RMSE 0.000341, 0.000312, and 0.000306. Table~\ref{tab:sparse-formulas} reports the recovered supports and evaluation metrics; additional sparse-formula details are reported in Appendix~\ref{app:sparse-formula-linear-baselines}. Thus, the KAN outputs admit a sparse signed Pauli summary matching the canonical GHZ relations. Because equivalent sparse descriptions can exist on the symmetric GHZ manifold, this result should be interpreted as recovery of a canonical support-and-sign convention rather than as proof of a unique formula.

\begin{table*}
\caption{\label{tab:sparse-formulas}
Sparse formula recovery results. Support is selected from all 63 Pauli channels and evaluated against the canonical target-specific GHZ supports. Precision and recall are computed separately for each target.}
\begin{ruledtabular}
\begin{tabular}{llcccc}
Target & Recovered support & Precision & Recall & $R^2$ & RMSE\\
\hline
$z$ &
$\mathrm{ZII}, \mathrm{IZI}, \mathrm{IIZ}, \mathrm{ZZZ}$ &
1.000 & 1.000 & 0.999999768 & 0.000341\\
$c$ &
$\mathrm{XXX}, \mathrm{XYY}, \mathrm{YXY}, \mathrm{YYX}$ &
1.000 & 1.000 & 0.999999602 & 0.000312\\
$s$ &
$\mathrm{YXX}, \mathrm{XYX}, \mathrm{XXY}, \mathrm{YYY}$ &
1.000 & 1.000 & 0.999999630 & 0.000306\\
\end{tabular}
\end{ruledtabular}
\end{table*}

\subsection{Linear sparse baselines}

We next ask whether Pauli-support recovery is specific to the KAN architecture, or whether it is already obtained by standard sparse linear regression on this near-linear GHZ benchmark. All four linear models were evaluated under the same default 1000-shot finite-shot condition as the KAN, following the protocol described in Methods. All four models achieved GHZ-subspace fidelity above $0.996$ and placed the canonical GHZ channels among their top-12 largest-magnitude coefficients (Table~\ref{tab:linear-baselines}). LASSO and elastic net recovered the canonical extended 12-channel GHZ support with mean top-12 extended-channel Jaccard $1.000\pm0.000$ and mean target-level support precision and recall both equal to 1.000. Their RMSE values were $0.013182 \pm 0.000299$ (LASSO) and $0.013224 \pm 0.000306$ (elastic net), compared with $0.012562$ for the default 1000-shot KAN.

OLS and ridge gave slightly higher fidelity than the sparse linear models, but their coefficient supports remained dense under finite-shot noise. As a result, they had low support precision ($\approx 0.07$) despite perfect recall. These results show that support recovery in this GHZ benchmark is not unique to the KAN: sparse linear models recover the same canonical Pauli support under the default finite-shot condition, whereas dense linear models provide accurate prediction without sparse support.

Noiseless and purely depolarized GHZ-family data further illustrate a support-identifiability caveat (see Appendix~\ref{app:sparse-formula-linear-baselines} for full results). In those settings, LASSO can select a single representative from a redundant GHZ-equivalent channel family, such as $z\approx\mathrm{ZII}$, rather than the symmetric four-term canonical formula.

The linear baselines therefore define the boundary of the KAN's contribution. Support recovery is not unique to the KAN: sparse linear regression already recovers the canonical Pauli support under this benchmark. The KAN contribution is instead neural pathway-level structural interpretability---the ability to inspect how measurement evidence is organized inside a pruned input-hidden-output graph before the final outputs are produced.

\begin{table*}
\caption{\label{tab:linear-baselines}
Linear sparse baselines at the default 1000-shot condition (GHZ-subspace fidelity and RMSE evaluated on the test set; results averaged over seeds 0--4). OLS and ridge are dense linear prediction references; LASSO and elastic net are sparse support-recovery references. For each model, the 12 largest-magnitude coefficients are identified by pooling across the three output-specific coefficient vectors and ranking by absolute magnitude. Top-12 Jaccard is computed between these coefficients and the canonical extended 12-channel GHZ support. Precision and recall are averaged over the target-specific supports for $z$, $c$, and $s$ using a coefficient threshold of $10^{-3}$.}
\begin{ruledtabular}
\begin{tabular}{lcccccc}
Model & Fidelity & RMSE & Top-12 Jaccard & Support precision & Support recall & Characteristic\\
\hline
OLS & $0.998050 \pm 0.000133$ & $0.012628 \pm 0.000315$ & $1.000 \pm 0.000$ & $0.071$ & $1.000$ & Dense coefficients\\
Ridge & $0.997866 \pm 0.000146$ & $0.012674 \pm 0.000315$ & $1.000 \pm 0.000$ & $0.074$ & $1.000$ & Dense coefficients\\
LASSO & $0.996252 \pm 0.000314$ & $0.013182 \pm 0.000299$ & $1.000 \pm 0.000$ & $1.000$ & $1.000$ & Sparse coefficients\\
Elastic net & $0.996165 \pm 0.000305$ & $0.013224 \pm 0.000306$ & $1.000 \pm 0.000$ & $1.000$ & $1.000$ & Sparse coefficients\\
\end{tabular}
\end{ruledtabular}
\end{table*}

\subsection{Controls, stress tests, and boundary conditions}

The preceding analyses indicate that the KAN selects GHZ-relevant Pauli channels, organizes them in a way consistent with the GHZ Pauli structure, and admits a signed sparse Pauli summary. We next test five possible failure modes: functional insufficiency, off-manifold boundary behavior, random-label artifacts, noise-level instability, and pruning-threshold sensitivity.

\begin{figure*}[!t]
\centering
\includegraphics[width=\textwidth]{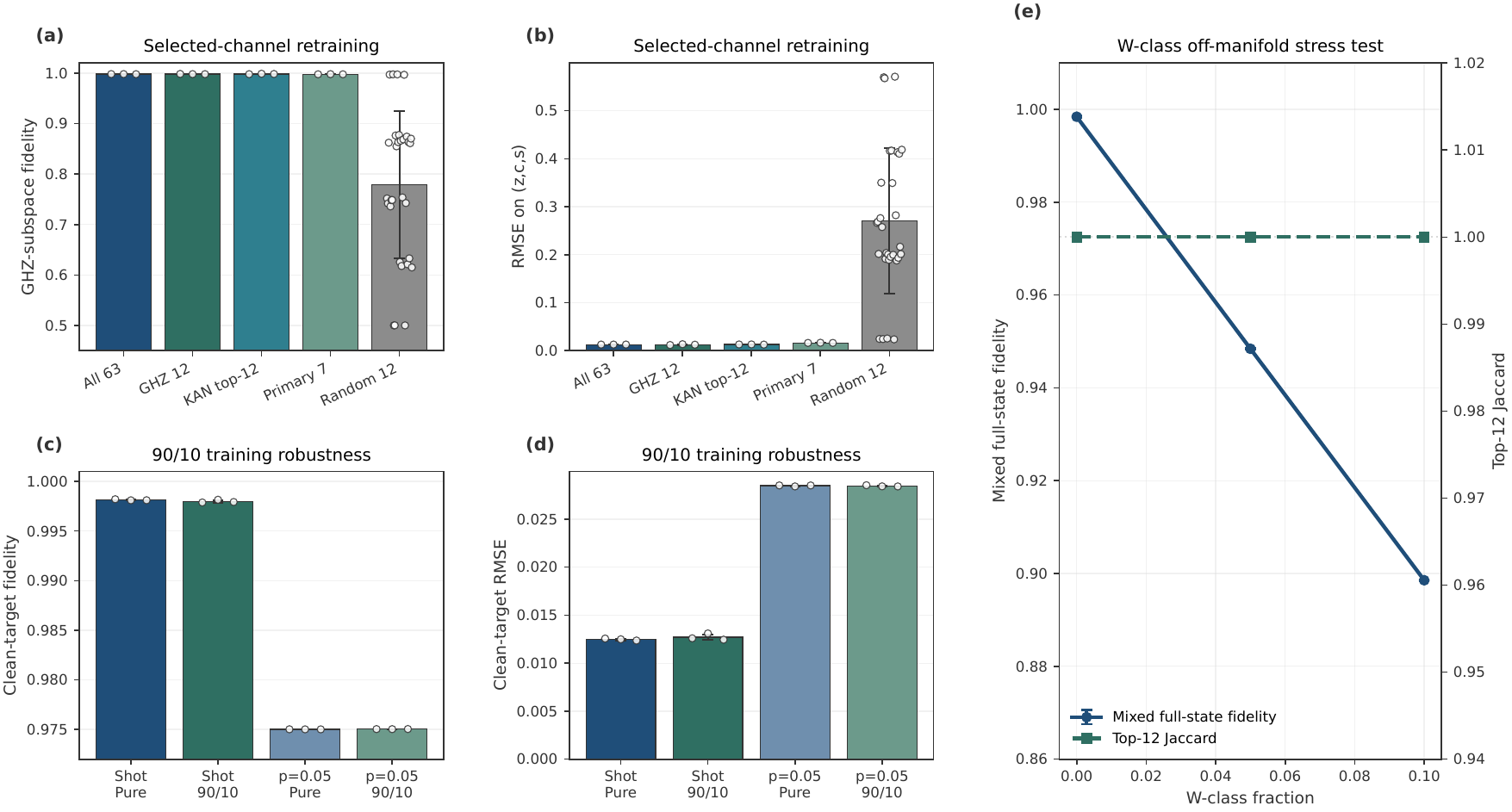}
\caption{\label{fig:selected-robustness}
Functional sufficiency and off-manifold boundary tests. All panels use the 1000-shot condition unless otherwise noted. (a,b) GHZ-subspace fidelity and RMSE for KANs retrained on five input subsets: all 63 channels, GHZ-theory 12, KAN-selected top-12, primary GHZ 7, and random 12-channel sets. Points show individual seed or repeat values; bars show mean $\pm$ standard deviation. (c,d) Clean-target fidelity and RMSE for pure-GHZ training versus 90/10 GHZ/other-state training, evaluated on pure-GHZ test states under 1000-shot finite-shot noise and depolarizing noise with $p=0.05$. (e) W-class off-manifold stress test: mixed full-state fidelity decreases as the W-class fraction increases, while the top-12 extended-channel Jaccard remains stable.}
\end{figure*}

\textbf{Functional sufficiency.} Selected-channel retraining tests whether the KAN-identified channels are sufficient when used as the only measurement inputs, rather than merely appearing important under an attribution score. KANs were retrained from scratch on five input configurations: all 63 Pauli channels (full-input reference), the 12-channel GHZ-theory set (analytic reference), the KAN-selected top-12 channels (learned subset), the 7-channel primary GHZ set (to probe redundancy in the extended representation), and random 12-channel subsets drawn with repeated random draws per seed across 30 total random-subset runs (negative control). At 1000 shots, KANs retrained on the KAN-selected top-12 channels achieved fidelity $0.998256 \pm 0.000404$, comparable to the all-63-channel reference ($0.997977 \pm 0.000131$) and the GHZ-theory 12-channel reference ($0.998067 \pm 0.000235$) (Table~\ref{tab:selected-retraining} and Fig.~\ref{fig:selected-robustness}a,b). The primary GHZ 7-channel set also retained high fidelity ($0.997761 \pm 0.000160$), although RMSE increased from $0.012782$ to $0.016318$. Random 12-channel subsets performed substantially worse (fidelity $0.778701 \pm 0.148555$). These results support functional sufficiency of the KAN-selected subset; they do not imply that the 12-channel set is uniquely minimal.

\begin{table*}
\caption{\label{tab:selected-retraining}
Selected-channel retraining. All retraining runs use the default 1000-shot finite-shot condition; random 12-channel subsets use 30 total runs (seeds 0--2, ten random draws per seed). GHZ-subspace fidelity and RMSE are evaluated on the test set.}
\begin{ruledtabular}
\begin{tabular}{lcccl}
Input subset & Channels & GHZ-subspace fidelity & RMSE & Characteristic\\
\hline
All 63 channels & 63 & $0.997977 \pm 0.000131$ & $0.012782 \pm 0.000016$ & Full-input reference\\
GHZ-theory 12 channels & 12 & $0.998067 \pm 0.000235$ & $0.012621 \pm 0.000848$ & Analytic support\\
KAN-selected top 12 channels & 12 & $0.998256 \pm 0.000404$ & $0.012928 \pm 0.000199$ & Learned support\\
Primary GHZ 7 channels & 7 & $0.997761 \pm 0.000160$ & $0.016318 \pm 0.000182$ & Redundant representation\\
Random 12 channels & 12 & $0.778701 \pm 0.148555$ & $0.270511 \pm 0.153834$ & Lower fidelity; higher RMSE variance\\
\end{tabular}
\end{ruledtabular}
\end{table*}

\textbf{Off-manifold boundary conditions.}
Two experiments probe the boundary of the GHZ-derived reconstruction rule. The 90/10 training experiment tests whether weak off-manifold contamination during training destabilizes the learned GHZ rule; validation and testing remain on pure GHZ-family states, so this is a weak training-contamination test rather than arbitrary-state tomography. This independent three-seed control uses a 400-sample pure-GHZ test set and is therefore distinct from the main 1000-shot noise sweep. At 1000 shots, pure-GHZ training achieved clean-target fidelity $0.998136 \pm 0.000059$ and mean RMSE $0.012470 \pm 0.000105$ on the pure-GHZ test set. Training with 10\% Haar-random other states achieved clean-target fidelity $0.997986 \pm 0.000147$ and mean RMSE $0.012713 \pm 0.000342$. The top-12 extended-channel Jaccard remained $1.000 \pm 0.000$ in both conditions. At depolarizing probability $p=0.05$, pure-GHZ training gave clean-target fidelity $0.975010 \pm 0.000003$ and RMSE $0.028479 \pm 0.000060$, while 90/10 training gave clean-target fidelity $0.975034 \pm 0.000006$ and RMSE $0.028454 \pm 0.000071$ (Fig.~\ref{fig:selected-robustness}c,d).

The W-class stress test probes controlled departure from the GHZ manifold. The external-ablation top-12 extended-channel Jaccard remained $1.000$, while the mixed full-state fidelity decreased from approximately $0.9984$ to $0.9484$ and $0.8986$ as the W-class fraction increased from $0$ to $0.05$ and $0.10$, respectively (Fig.~\ref{fig:selected-robustness}e). This defines an off-manifold boundary rather than a general arbitrary-state tomography result: the model continues to apply the GHZ-derived channel-selection rule outside its training manifold, while the resulting GHZ-block prediction no longer represents the full off-manifold state.

\begin{figure*}[!t]
\centering
\includegraphics[width=\textwidth]{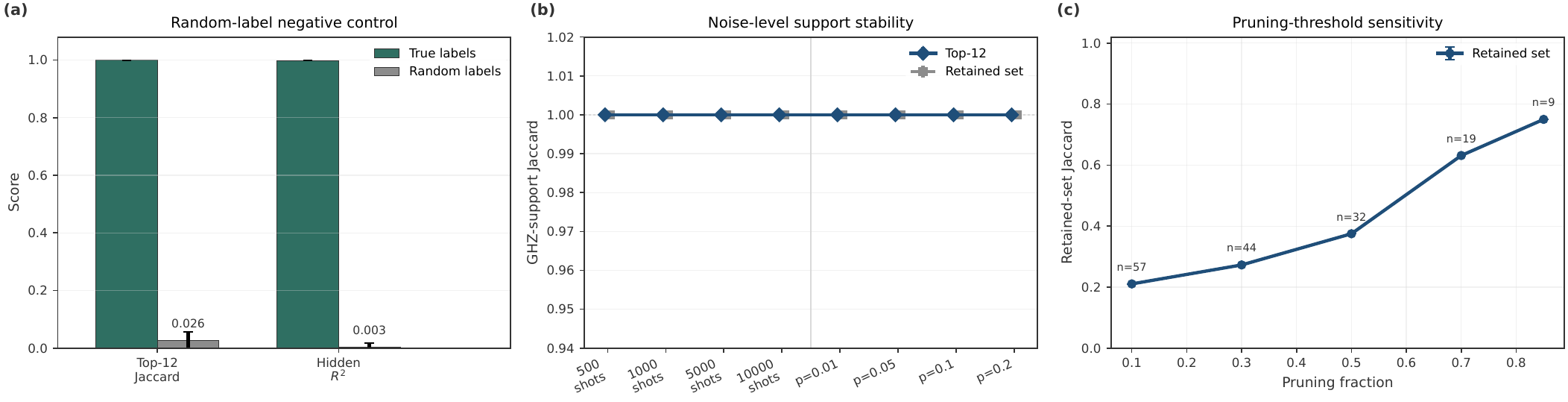}
\caption{\label{fig:control-stability}
Negative controls and support-stability analyses. All panels use seeds 0--4 unless otherwise noted. (a) Random-label negative control: shuffling the training input--target pairing collapses both top-12 extended-channel recovery and hidden-subspace alignment. (b) Noise-level support stability: across the tested finite-shot and depolarizing noise ranges, the top-12 extended-channel Jaccard and retained-set Jaccard remain at unity under the default support-recovery and pruning protocol. (c) Pruning-threshold sensitivity: increasing sparsification changes the retained-set size and enriches the retained input set for GHZ-relevant Pauli channels. The retained-set Jaccard values are interpreted relative to the ceiling imposed by the number of retained channels.}
\end{figure*}

\textbf{Random-label negative control.}
Random-label training tests whether the support and hidden-alignment signals survive when the input--target relation is destroyed. When the training input--target pairing is shuffled, while evaluation remains on unshuffled held-out data, the top-12 extended-channel Jaccard drops from $1.000\pm0.000$ to $0.026\pm0.030$, and the hidden-subspace probe $R^2$ drops from $0.9985\pm0.0007$ to $0.0034\pm0.0129$ (Fig.~\ref{fig:control-stability}a). The collapse of both metrics under random labeling indicates that the KAN's channel preference and internal GHZ-aligned representation depend on the physical input--target relation rather than on dataset-specific shortcuts.

\textbf{Noise-level support stability.}
Across the tested finite-shot (500--10000 shots) and depolarizing ($p=0.01$--$0.20$) noise sweeps, the top-12 extended-channel Jaccard remained $1.000\pm0.000$ under the default support-recovery protocol (Fig.~\ref{fig:control-stability}b). The depolarizing sweep should be interpreted as a consistency check, because this noise model preserves exact zero expectations on non-GHZ Pauli channels, whereas finite-shot sampling perturbs every measured channel. Thus, while randomization eliminates the input-support pattern, varying the noise level within the tested ranges does not. An extended 10-seed analysis further confirms that support recovery is stable across random initializations: the KAN retained-set and external-ablation top-12 extended-channel Jaccard values remained $1.000\pm0.000$ across seeds, and the 20-seed hidden-subspace analysis at 1000 shots gives a held-out probe score of $R^2=0.9985\pm0.0007$, consistent with the results reported in Sec.~III.C.

\textbf{Pruning-threshold sensitivity.}
The pruning-threshold sweep tests sensitivity to the sparsity level (Fig.~\ref{fig:control-stability}c). Across five seeds, the retained count decreased from 57 at prune fraction 0.1 to 9 at prune fraction 0.85, while the retained-set Jaccard increased from $0.211 \pm 0.000$ to $0.750 \pm 0.000$. Because retained-set Jaccard is bounded by retained-set size, its absolute value should not be interpreted as a direct support-recovery score. For example, when 57 channels are retained, the maximum possible Jaccard with a 12-channel reference support is $12/57=0.211$; when only 9 channels are retained, the maximum possible value is $9/12=0.750$. The observed values match these size-imposed ceilings, indicating that pruning progressively enriches the retained set for GHZ-relevant channels. This analysis is therefore a sparsity-enrichment sensitivity check rather than the primary support-recovery metric.

Table~\ref{tab:S3} provides a full tabular summary of the control analyses. A compact MLP comparison is reported in Appendix~\ref{app:kan-mlp-comparison} and is referenced below as an architecture-level reference.

\section{Discussion}

\subsection{What the probes demonstrate}

The interpretation of the trained KAN should be viewed as a consistency chain rather than as the outcome of a single attribution method. External ablation identifies the Pauli channels on which the trained model functionally depends. Pathway analysis then asks whether those channels are organized inside the sparse KAN in a way that is consistent with the population and off-diagonal Pauli structure. Hidden-subspace probes test whether the target variables are linearly accessible from the internal activations. Sparse formula recovery provides a signed post-hoc summary of the learned input-output rule. Selected-channel retraining tests functional sufficiency of the recovered support, while shuffled-input, decoy-channel, random-label, noise-stability, and pruning-threshold controls test whether the interpretation survives common artifact checks. The resulting evidence chain is stronger than a single feature-importance or visualization analysis alone~\cite{Murdoch2019,Roscher2020}, but it should still be interpreted as convergent structural evidence rather than as a causal proof of the learned computation.

The sparse linear baselines define the central boundary of the claim. The GHZ-subspace task is deliberately near-linear, and LASSO and elastic net recover the canonical Pauli support with the same top-12 extended-channel Jaccard of $1.000 \pm 0.000$ as the KAN, with mean target-level support precision and recall both equal to 1.000, under the default 1000-shot finite-shot condition. Thus, channel selection and sparse formula recovery alone are not sufficient reasons to prefer a KAN over sparse linear regression. External ablation and sparse formula recovery are architecture-agnostic analyses, and hidden-representation probes can be applied to any neural model with internal activations.

Conceptually, this work proposes a reorientation of neural quantum state tomography: from treating reconstruction error as the sole endpoint to auditing the learned reconstruction rule against known physical structure. The sparsified KAN is a concrete instantiation of this inspection-oriented paradigm, because its pruned edge structure makes channel support, internal routing, hidden alignment, and signed formula consistency measurable within one workflow. By making the learned rule itself an object of evaluation, this opens a direction beyond conventional fidelity optimization. More broadly, structured neural QST studies should report fidelity together with inspection metrics such as support overlap, hidden-subspace alignment, and formula consistency whenever a validated physical reference is available.

The distinction is therefore not between an interpretable KAN and an uninterpretable sparse linear model. Sparse linear regression gives formula-level interpretability: a coefficient vector states the final sparse linear rule. What it does not provide is a pathway-level neural structural object showing how a neural reconstruction model organizes measurement evidence internally before producing the final outputs. In this benchmark, the KAN contribution is precisely this pathway-level structural interpretability inside a neural reconstruction model, complementing rather than replacing the formula-level reference provided by sparse linear regression.

When a sparse linear coefficient vector provides the required predictive and support-recovery performance, sparse linear regression is the more direct and parsimonious choice. Neural reconstruction models may become more useful when the learned map involves raw-count inputs, calibration effects, unknown measurement distortions, missing-measurement inference, or nonlinear parameter outputs such as $\theta$ and $\phi$. In such settings, KAN-style edge and pathway objects could help inspect whether physically meaningful measurement structure remains visible inside the learned model. This is the methodological role of the KAN here: it provides inspectable neural structure, not superior sparse regression on the present near-linear benchmark.

\begin{figure*}[!t]
\centering
\includegraphics[width=\textwidth]{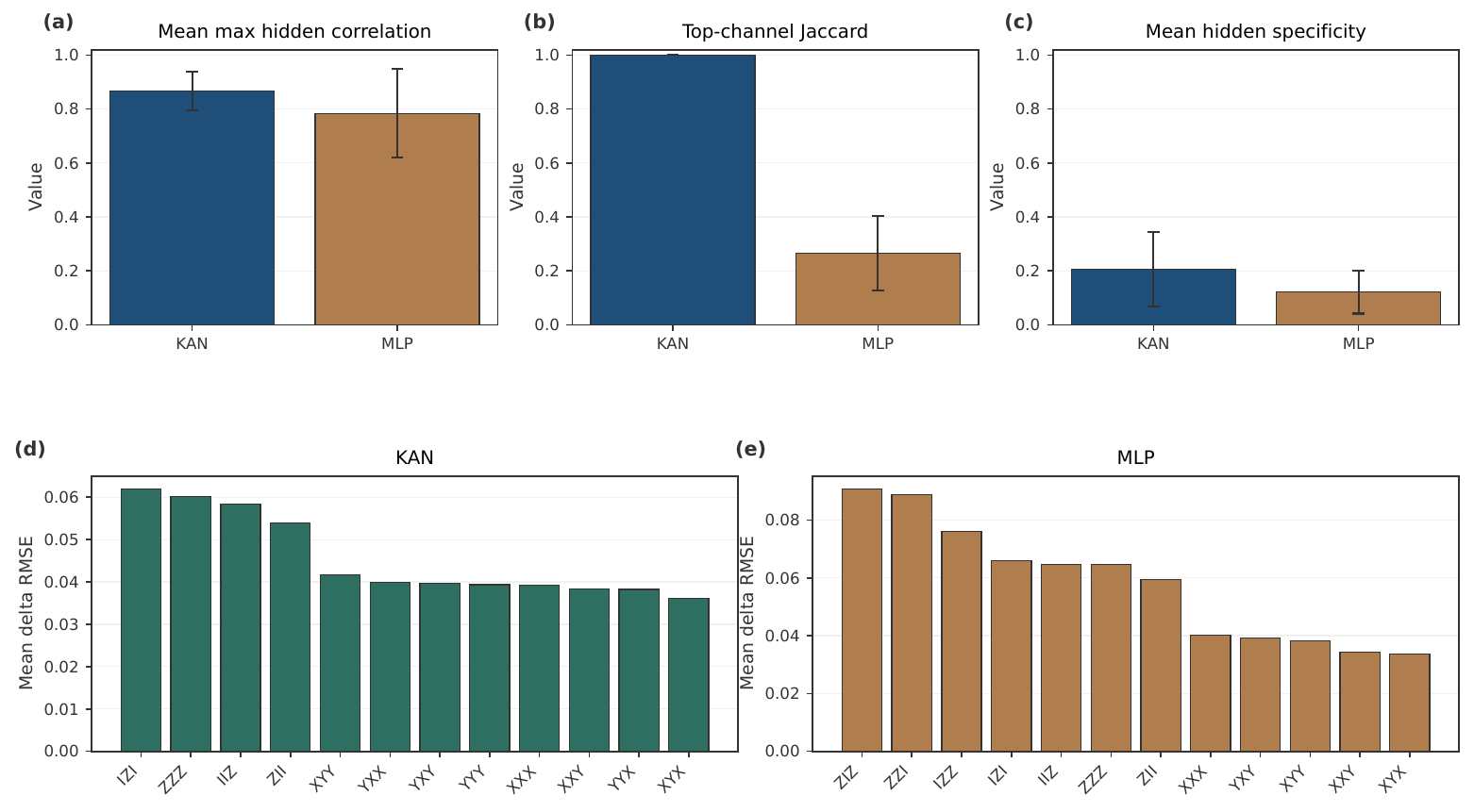}
\caption{\label{fig:mlp-comparison}
Architecture-level reference comparison between the sparsified KAN and a width-matched compact MLP. All metrics are computed at the default 1000-shot condition (seeds 0--4). (a) Mean maximum absolute hidden-unit correlation per GHZ variable. (b) Top-channel Jaccard. (c) Mean hidden-unit specificity. (d,e) External-ablation channel rankings for the KAN and MLP, reported as mean $\Delta$RMSE after zeroing each input channel. Full details are provided in Appendix~\ref{app:kan-mlp-comparison}. The figure provides supporting context for the pathway-level interpretability discussion and is not an exhaustive architecture benchmark.
}
\end{figure*}

The compact MLP comparison provides supporting architecture-level context for this distinction (Fig.~\ref{fig:mlp-comparison} and Appendix~\ref{app:kan-mlp-comparison}). For the MLP, the internal support score is defined through a post-hoc input-hidden attribution proxy rather than a native pruned pathway object. The MLP can be probed with external ablation and hidden-representation statistics, but it does not natively provide KAN-style edge functions or pruned pathways linking Pauli observables, hidden units, and reconstructed variables. We therefore treat the MLP comparison only as supporting context for the distinction between predictive accuracy and pathway-level structural interpretability. The post-hoc proxy enables a top-channel ranking, but it is not equivalent to a native pruned pathway representation. A fair architecture comparison would require a dedicated study of interpretability-preserving regularizers or alternative structured MLP designs. Our aim here is to show that the KAN offers directly inspectable pathway objects, not to prove that its hidden representations are universally more interpretable. Thus, the MLP results are not claimed as standalone evidence of KAN superiority.

\subsection{Scope and limitations}

The scope of the study is deliberately limited. The task is three-qubit GHZ-subspace tomography, and the model outputs are $(z,c,s)$, not a complete $8 \times 8$ density matrix. The present implementation also assumes Pauli-expectation inputs rather than raw measurement counts. Because the GHZ family has compact, known Pauli structure, it serves as a controlled analytic testbed for model inspection rather than as a representative arbitrary-state tomography benchmark. The noise models are controlled simulations of finite-shot sampling and depolarizing attenuation, not calibrated experimental noise processes. We therefore view the present study as a controlled demonstration of an inspection-oriented QST workflow, not as a general-purpose tomography algorithm.

The off-manifold experiments have similar limits. The 90/10 GHZ/other-state training condition tests weak off-manifold contamination during training while validation and testing remain on pure GHZ-family data. The W-class stress test probes controlled departure from the GHZ manifold, not arbitrary-state tomography. It should be interpreted as a boundary test: the model continues to apply a GHZ-derived channel-selection rule to non-GHZ inputs, but it was not trained or parameterized to represent the W-class subspace. Thus, the stress test identifies where the GHZ-derived reconstruction rule ceases to describe the full state, rather than establishing robustness for arbitrary three-qubit states.

Sparse formula recovery should also be interpreted with caution. On a highly symmetric manifold, multiple Pauli expressions can be redundant or statistically equivalent, and sparse recovery depends on a support-selection convention. The recovered support matches a canonical GHZ Pauli representation rather than establishing a unique underlying physical rule. Similarly, the compact MLP comparison addresses structural interpretability in this setting, not exhaustive benchmarking of neural tomography architectures. The hidden-subspace comparison with the compact MLP is also confounded by predictive performance; we treat it as supporting context rather than standalone evidence that KAN hidden representations are intrinsically more interpretable.

\subsection{Outlook}

Future work should test whether the same inspection strategy extends to settings where the reconstruction rule is less analytically transparent, including four- and five-qubit GHZ states, W states, cluster states, low-rank mixed states, unknown noise models, adaptive measurement design, and real experimental count data. These settings would test a different question from the present GHZ benchmark: not whether a KAN can rediscover a known sparse Pauli rule, but whether pathway-level inspection remains useful when the relevant measurement structure is only partially known, distorted by calibration effects, or learned from incomplete data. We are also extending the framework to larger-scale simulations of eight-qubit Haar-random pure states under sparse measurements. This setting will test whether KAN-based inspection remains informative when measurements are incomplete and no compact analytic Pauli rule is available.

In such applications, sparsified KANs may be useful as physics-guided inspection tools. They can help identify which measurement channels a reconstruction model uses, how those channels are organized inside the model, and whether the learned rule remains compatible with known physical constraints. This outlook is methodological rather than promotional: the present results motivate applying the inspection workflow to more complex quantum systems, while the evidence in this paper remains anchored in a controlled GHZ-family testbed.

\section{Conclusion}

We introduced a sparsified KAN framework for interpretable GHZ-subspace tomography. The framework treats the trained model not only as a reconstruction map from Pauli expectation values to the variables $(z,c,s)$, but also as an inspectable neural object whose retained inputs, hidden units, pathways, and output-level summaries can be compared with known Pauli structure. The three-qubit GHZ family provides a controlled analytic setting for this comparison, rather than a general-purpose benchmark for arbitrary three-qubit tomography.

Within this setting, the KAN reconstructs the GHZ population and off-diagonal variables under finite-shot and depolarizing noise, identifies the expected GHZ-relevant Pauli channels from the full 63-channel input, and exhibits pathway-level organization consistent with the population and coherence structure of the benchmark. Hidden-subspace probes, sparse formula recovery, selected-channel retraining, and control analyses further support the interpretation that the learned rule reflects the analytic GHZ Pauli structure rather than column-order artifacts, irrelevant-channel shortcuts, or random-label effects.

The linear sparse baselines define the central boundary of the result. On this near-linear benchmark, LASSO and elastic net recover the same Pauli support with comparable reconstruction accuracy. Thus, the contribution of the KAN is not superior sparse regression. Its contribution is pathway-level structural interpretability inside a neural reconstruction model: sparse linear regression provides a formula-level reference, whereas the sparsified KAN provides inspectable neural structure for examining how measurement information is organized inside the model.

The present evidence remains anchored in a controlled three-qubit GHZ-subspace setting and does not establish a general solution to arbitrary-state quantum tomography. Rather, the study provides a methodological step toward inspectable neural reconstruction in more complex quantum systems, where learned models may be useful only if their internal organization can be audited against physical structure and known constraints. Accordingly, we recommend pairing fidelity with inspection metrics as a standard practice for structured tomography tasks whenever reference structure is available. This shift from prediction-only optimization to structured auditing opens a direction beyond conventional fidelity optimization for interpretable quantum state learning, in which models are evaluated not only by what they predict but also by whether their internal organization respects physical constraints.
\section*{Data and Code Availability}

The code used to generate the datasets, train the models, compute the inspection metrics, and reproduce the figures will be made available in a public repository upon publication. The repository will include the scripts for GHZ-family data generation, Pauli-expectation construction, KAN and baseline training, noise simulations, pruning and pathway analysis, sparse formula recovery, selected-channel retraining, and the control experiments reported in the main text and the appendices. The numerical data underlying the figures and tables will be provided in machine-readable form together with the repository. No human-subject, clinical, or restricted-access experimental data are used in this study.

\begin{acknowledgments}
This work is supported by the National Key Research and Development Program of China (2024YFB4504103). This work is also supported by Jiangsu Province Engineering Research Center of IntelliSense Technology and System.
\end{acknowledgments}

\appendix

\setcounter{table}{0}
\setcounter{figure}{0}
\renewcommand{\thetable}{S\arabic{table}}
\renewcommand{\thefigure}{S\arabic{figure}}
\section{GHZ benchmark implementation}\label{app:ghz-benchmark}

This note specifies the benchmark implementation used to generate the
GHZ-family data and Pauli-expectation inputs.

\subsection{Complete 63-channel Pauli ordering}

Table~\ref{tab:S1} gives the complete input-channel ordering used throughout the
study. Each Pauli string follows the tensor-product convention
$\sigma_i \otimes \sigma_j \otimes \sigma_k$. Characters are read from
left to right, with the $k$-th character acting on the $k$-th qubit.

\begin{table*}[t]
\caption{\label{tab:S1}
Complete ordering of the 63 non-identity three-qubit Pauli expectation
values used as input channels. The first 12 entries are the extended
GHZ-relevant support used for overlap metrics. The remaining entries
follow the implementation ordering after these 12 channels are removed.
}
\begin{ruledtabular}
\begin{tabular}{@{}c@{\quad}l@{\hspace{0.9cm}\vline\hspace{0.9cm}}c@{\quad}l@{\hspace{0.9cm}\vline\hspace{0.9cm}}c@{\quad}l@{}}
Channel & Pauli string & Channel & Pauli string & Channel & Pauli string\\
\hline
 1 & ZII & 22 & IYZ & 43 & YXZ\\
 2 & IZI & 23 & IZX & 44 & YYI\\
 3 & IIZ & 24 & IZY & 45 & YYZ\\
 4 & ZZZ & 25 & IZZ & 46 & YZI\\
 5 & XXX & 26 & XII & 47 & YZX\\
 6 & YXX & 27 & XIX & 48 & YZY\\
 7 & XYX & 28 & XIY & 49 & YZZ\\
 8 & XXY & 29 & XIZ & 50 & ZIX\\
 9 & XYY & 30 & XXI & 51 & ZIY\\
10 & YXY & 31 & XXZ & 52 & ZIZ\\
11 & YYX & 32 & XYI & 53 & ZXI\\
12 & YYY & 33 & XYZ & 54 & ZXX\\
13 & IIX & 34 & XZI & 55 & ZXY\\
14 & IIY & 35 & XZX & 56 & ZXZ\\
15 & IXI & 36 & XZY & 57 & ZYI\\
16 & IXX & 37 & XZZ & 58 & ZYX\\
17 & IXY & 38 & YII & 59 & ZYY\\
18 & IXZ & 39 & YIX & 60 & ZYZ\\
19 & IYI & 40 & YIY & 61 & ZZI\\
20 & IYX & 41 & YIZ & 62 & ZZX\\
21 & IYY & 42 & YXI & 63 & ZZY\\
\end{tabular}
\end{ruledtabular}
\end{table*}

\subsection{Data-generation procedure}

Algorithm~S1 summarizes the data-generation procedure used for the
GHZ-family experiments. The variables $(z,c,s)$ are extracted from the
$\ket{000}$ and $\ket{111}$ amplitudes as defined in Sec.~II.B of the
main text.

\medskip
\noindent\textbf{Algorithm S1:} GHZ-family data generation.
\begin{enumerate}
    \item Sample $\theta \sim \mathrm{Uniform}(0,\pi)$,
          $\phi \sim \mathrm{Uniform}(-\pi,\pi)$.
    \item Construct the pure state
          $\ket{\psi} = \cos(\theta/2)\ket{000}
                      + e^{i\phi}\sin(\theta/2)\ket{111}$.
    \item Compute the $8\times 8$ density matrix
          $\rho = \ket{\psi}\bra{\psi}$.
    \item Evaluate the 63 Pauli expectation values
          $\langle P_i\rangle = \mathrm{Tr}(\rho P_i)$
          for $i = 1,\dots,63$ in the order of Table~\ref{tab:S1}.
    \item Extract $z = \rho_{000,000} - \rho_{111,111}$,
          $c = 2\,\mathrm{Re}(\rho_{000,111})$,
          $s = 2\,\mathrm{Im}(\rho_{000,111})$.
    \item Apply noise as specified in Sec.~II.C of the main text
          (finite-shot sampling or depolarizing attenuation).
\end{enumerate}

The default split contains 1400 training samples, 300 validation samples,
and 300 test samples. Noise sweeps use seeds 0--4, and extended controls
use seeds 0--9 unless noted otherwise. The 20-seed hidden-subspace
analysis is performed at the 1000-shot condition. The sign convention for
$s$ and the fidelity definition follow Secs.~II.B and II.E of the main
text.

\section{Model training and retraining protocols}\label{app:training-protocols}

This note records the model architectures, training schedule, data splits,
and restricted-input retraining settings needed to reproduce the reported
experiments.

\subsection{KAN hyperparameters and training}

\begin{table}[H]
\caption{\label{tab:S2_kan}
KAN architecture and training hyperparameters.
}
\begin{ruledtabular}
\begin{tabular}{ll}
Parameter & Value\\
\hline
Input dimension & 63\\
Hidden width & 6\\
Output dimension & 3\\
Grid size & 5\\
Spline order & 3\\
Optimizer (stage 1) & Adam\\
Learning rate (Adam) & 0.01\\
Pruning threshold (default) & 0.01\\
Optimizer (stage 2) & LBFGS\\
Loss function & MSE on $(z,c,s)$\\
\end{tabular}
\end{ruledtabular}
\end{table}

Training is performed in two stages. Stage~1 uses Adam optimization for
200 epochs, followed by pruning of nodes and edges whose pykan importance
scores fall below the pruning threshold. The safe-pruning routine keeps at
least one hidden node when thresholding would otherwise remove the
intermediate layer. After pruning, input indices are mapped back to the
original Pauli labels. Stage~2 fine-tunes the pruned model with LBFGS for
100 iterations. The resulting model is used for test-set evaluation and
all post-training inspection analyses.

\subsection{MLP baseline hyperparameters}

The width-matched MLP baseline settings are summarized in Table~\ref{tab:S2_mlp}.

\begin{table}[H]
\caption{\label{tab:S2_mlp}
Width-matched bottleneck MLP hyperparameters.
}
\begin{ruledtabular}
\begin{tabular}{ll}
Parameter & Value\\
\hline
Input dimension & 63\\
Hidden width & 6\\
Output dimension & 3\\
Activation & ReLU\\
Optimizer & Adam\\
Learning rate & 0.01\\
Training epochs & 300\\
Loss function & MSE on $(z,c,s)$\\
\end{tabular}
\end{ruledtabular}
\end{table}

\subsection{Data splits and seed conventions}

The data splits and seed ranges used across the control and stability analyses are summarized in Table~\ref{tab:S2_splits}.

\begin{table}[H]
\caption{\label{tab:S2_splits}
Data splits and seed conventions used across experiments.
}
\begin{ruledtabular}
\begin{tabular}{lcl}
Experiment & Seeds & Notes\\
\hline
Main noise sweeps & 0--4 & 1400/300/300 split\\
Extended controls & 0--9 & Additional 5 seeds\\
Hidden-subspace alignment & 0--19 & 20 seeds, 1000-shot\\
Selected-channel retraining & 0--2 & 10 random draws per seed\\
Shuffled-input retraining & 0--2 & 3 retraining runs\\
\end{tabular}
\end{ruledtabular}
\end{table}

\subsection{Selected-channel retraining configurations}

Table~\ref{tab:S2_retrain} defines the five input configurations used for selected-channel
retraining in Sec.~III.E of the main text.

\begin{table}[H]
\caption{\label{tab:S2_retrain}
Input configurations for selected-channel retraining.
}
\begin{ruledtabular}
\begin{tabular}{lcl}
Configuration & Channels & Description\\
\hline
All 63 & 63 & Full-input reference\\
GHZ-theory 12 & 12 & $\mathcal{P}_{\mathrm{GHZ}}^{(12)}$\\
KAN-selected top 12 & 12 & Top-12 ablation ranking\\
Primary GHZ 7 & 7 & $\mathcal{P}_{\mathrm{GHZ}}^{(7)}$\\
Random 12 & 12 & 30 runs, 10 draws/seed over seeds 0--2\\
\end{tabular}
\end{ruledtabular}
\end{table}

\section{Extended control-experiment diagnostics}\label{app:control-diagnostics}

This note reports the extended diagnostics supporting the robustness and
control analyses in Figs.~5 and~6 of the main text.

\subsection{Per-noise-level support stability}

Table~\ref{tab:S3_noise} summarizes the top-12 extended-channel Jaccard and retained-set
Jaccard across the tested finite-shot and depolarizing noise levels.
Both metrics remain $1.000 \pm 0.000$ in this sweep, indicating stable
recovery of the GHZ Pauli support within the tested noise range.

\begin{table*}[t]
\caption{\label{tab:S3_noise}
Per-noise-level support-stability metrics (3 seeds per condition).
}
\begin{ruledtabular}
\begin{tabular}{lccccc}
Noise type & Value & Seeds & Top-12 Jaccard & Retained-set Jaccard\\
\hline
Finite-shot & \phantom{0}500 shots & 3 & $1.000 \pm 0.000$ & $1.000 \pm 0.000$\\
Finite-shot & 1000 shots & 3 & $1.000 \pm 0.000$ & $1.000 \pm 0.000$\\
Finite-shot & 5000 shots & 3 & $1.000 \pm 0.000$ & $1.000 \pm 0.000$\\
Finite-shot & 10000 shots & 3 & $1.000 \pm 0.000$ & $1.000 \pm 0.000$\\
Depolarizing & $p = 0.01$ & 3 & $1.000 \pm 0.000$ & $1.000 \pm 0.000$\\
Depolarizing & $p = 0.05$ & 3 & $1.000 \pm 0.000$ & $1.000 \pm 0.000$\\
Depolarizing & $p = 0.10$ & 3 & $1.000 \pm 0.000$ & $1.000 \pm 0.000$\\
Depolarizing & $p = 0.20$ & 3 & $1.000 \pm 0.000$ & $1.000 \pm 0.000$\\
\end{tabular}
\end{ruledtabular}
\end{table*}

\subsection{Per-seed random-label negative control}

Table~\ref{tab:S3_random} gives the per-seed metrics for the random-label negative
control. With true labels, the top-12 Jaccard and hidden-probe $R^2$ are
near ceiling. After label randomization, both metrics collapse toward
chance-level values, showing that the observed channel preference depends
on the physical input--target relation.

\begin{table*}[t]
\caption{\label{tab:S3_random}
Random-label negative-control metrics (10 seeds, 1000-shot). The
true-label row gives the corresponding 10-seed reference summary.
}
\begin{ruledtabular}
\begin{tabular}{ccc}
Seed & Top-12 Jaccard & Probe $R^2_{z,c,s}$\\
\hline
0 & 0.000 & -0.0127\\
1 & 0.000 & -0.0012\\
2 & 0.000 & -0.0004\\
3 & 0.043 & 0.0276\\
4 & 0.000 & -0.0003\\
5 & 0.091 & -0.0001\\
6 & 0.000 & -0.0041\\
7 & 0.043 & 0.0140\\
8 & 0.043 & -0.0113\\
9 & 0.043 & 0.0230\\
\hline
Random-label mean $\pm$ SD & $0.026 \pm 0.030$ & $0.0034 \pm 0.0129$\\
True-label mean $\pm$ SD & $1.000 \pm 0.000$ & $0.9985 \pm 0.0007$\\
\end{tabular}
\end{ruledtabular}
\end{table*}

\subsection{Random-label importance heatmap}

\begin{figure*}[t]
\centering
\includegraphics[width=0.85\textwidth]{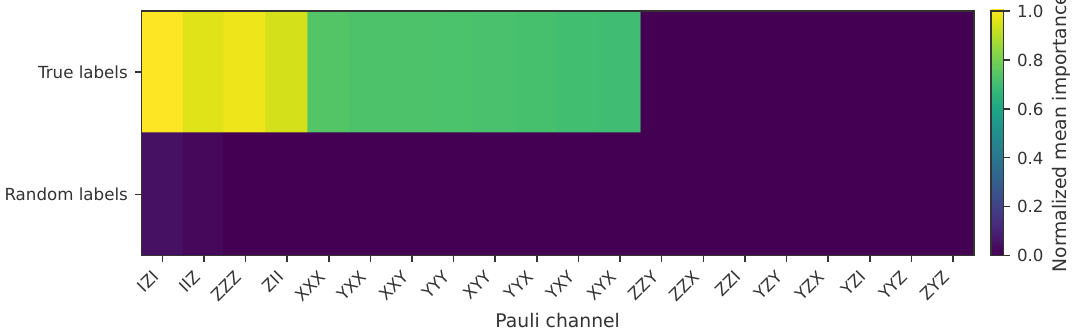}
\caption{\label{fig:S1}
Random-label external-ablation importance heatmap.
}
\end{figure*}

Figure~\ref{fig:S1} compares external-ablation importance across all 63 Pauli
channels for true-label and random-label training. With true labels, the
12 GHZ-relevant channels dominate the ranking. With randomized labels,
importance becomes low and unstructured.

\subsection{Random-initialization robustness}

Table~\ref{tab:S3_seed} reports the 10-seed support-stability analysis at the 1000-shot
condition. Both the retained-set and external-ablation top-12 Jaccard
values remain $1.000 \pm 0.000$ across seeds.

\begin{table*}[t]
\caption{\label{tab:S3_seed}
10-seed support-stability summary (1000-shot).
}
\begin{ruledtabular}
\begin{tabular}{cccc}
Seed & Retained-set Jaccard & Ablation top-12 Jaccard & Fidelity\\
\hline
0 & 1.000 & 1.000 & 0.9984\\
1 & 1.000 & 1.000 & 0.9983\\
2 & 1.000 & 1.000 & 0.9981\\
3 & 1.000 & 1.000 & 0.9982\\
4 & 1.000 & 1.000 & 0.9980\\
5 & 1.000 & 1.000 & 0.9979\\
6 & 1.000 & 1.000 & 0.9981\\
7 & 1.000 & 1.000 & 0.9982\\
8 & 1.000 & 1.000 & 0.9979\\
9 & 1.000 & 1.000 & 0.9982\\
\hline
Mean $\pm$ SD & $1.000 \pm 0.000$ & $1.000 \pm 0.000$ & $0.9981 \pm 0.0002$\\
\end{tabular}
\end{ruledtabular}
\end{table*}

\subsection{20-seed hidden-subspace alignment}

Table~\ref{tab:S2} reports the 20-seed hidden-subspace alignment statistics
referenced in the main text. The metrics include maximum absolute
hidden-unit correlations for each GHZ variable, hidden linear-probe
$R^2$ values, and CCA canonical correlations.

\begin{table}[H]
\caption{\label{tab:S2}
20-seed hidden-subspace alignment statistics (1000-shot). Values are
mean $\pm$ standard deviation across seeds 0--19.
}
\begin{ruledtabular}
\begin{tabular}{lc}
Metric & Mean $\pm$ SD\\
\hline
$\max|\mathrm{corr}(h,z)|$ & $0.9713 \pm 0.0518$\\
$\max|\mathrm{corr}(h,c)|$ & $0.8579 \pm 0.1551$\\
$\max|\mathrm{corr}(h,s)|$ & $0.8456 \pm 0.1648$\\
Mean max $|\mathrm{corr}|$ & $0.8916 \pm 0.0613$\\
Probe $R^2_z$ & $0.9988 \pm 0.0006$\\
Probe $R^2_c$ & $0.9979 \pm 0.0021$\\
Probe $R^2_s$ & $0.9985 \pm 0.0009$\\
Probe $R^2_{z,c,s}$ & $0.9985 \pm 0.0007$\\
Probe $R^2_{c,s}$ & $0.9982 \pm 0.0010$\\
CCA $(z,c,s)$ component 1 & $0.9997 \pm 0.0001$\\
CCA $(z,c,s)$ component 2 & $0.9994 \pm 0.0001$\\
CCA $(z,c,s)$ component 3 & $0.9985 \pm 0.0011$\\
CCA $(c,s)$ component 1 & $0.9995 \pm 0.0001$\\
CCA $(c,s)$ component 2 & $0.9987 \pm 0.0010$\\
\end{tabular}
\end{ruledtabular}
\end{table}

\subsection{Pruning-threshold sensitivity}

Table~\ref{tab:S3_prune} reports the pruning-threshold sensitivity analysis shown in
Fig.~6c of the main text. As the prune fraction increases, the retained
channel count decreases and the retained-set Jaccard with the GHZ
reference support increases. Because this Jaccard value is bounded by the
retained-set size, it is interpreted relative to the size-imposed ceiling.
The observed values therefore provide an enrichment diagnostic rather than
a stand-alone support-recovery proof: stronger pruning removes many
irrelevant channels, while the selected-channel retraining and external
ablation analyses provide the primary evidence for recovery of the GHZ
support.

\begin{table*}[t]
\caption{\label{tab:S3_prune}
Pruning-threshold sensitivity (5 seeds, 1000-shot).
}
\begin{ruledtabular}
\begin{tabular}{cccc}
Prune frac. & Retained count & Retained-set Jaccard & Notes\\
\hline
0.10 & $57.0 \pm 0.0$   & $0.211 \pm 0.000$ & See Fig.~6c\\
0.30 & $44.0 \pm 0.0$   & $0.273 \pm 0.000$ & See Fig.~6c\\
0.50 & $32.0 \pm 0.0$   & $0.375 \pm 0.000$ & See Fig.~6c\\
0.70 & $19.0 \pm 0.0$   & $0.632 \pm 0.000$ & See Fig.~6c\\
0.85 & $\phantom{0}9.0 \pm 0.0$   & $0.750 \pm 0.000$ & See Fig.~6c\\
\end{tabular}
\end{ruledtabular}
\end{table*}

\subsection{Extended 90/10 training-contamination metrics}

Table~\ref{tab:S3_90} reports the per-condition metrics for the 90/10 off-manifold
training experiment.

\begin{table*}[t]
\caption{\label{tab:S3_90}
90/10 training-contamination metrics (3 seeds). All metrics evaluated
on pure-GHZ test states.
}
\begin{ruledtabular}
\begin{tabular}{lccc}
Condition & Fidelity & RMSE & Top-12 Jaccard\\
\hline
Pure GHZ, 1000-shot    & $0.998136 \pm 0.000059$ & $0.012470 \pm 0.000105$ & $1.000 \pm 0.000$\\
90/10, 1000-shot       & $0.997986 \pm 0.000147$ & $0.012713 \pm 0.000342$ & $1.000 \pm 0.000$\\
Pure GHZ, $p=0.05$    & $0.975010 \pm 0.000003$ & $0.028479 \pm 0.000060$ & $1.000 \pm 0.000$\\
90/10, $p=0.05$       & $0.975034 \pm 0.000006$ & $0.028454 \pm 0.000071$ & $1.000 \pm 0.000$\\
\end{tabular}
\end{ruledtabular}
\end{table*}

\subsection{W-class off-manifold stress-test metrics}

Table~\ref{tab:S3_w} reports the W-class off-manifold stress test shown in
Fig.~5e of the main text. Mixed full-state fidelity decreases as the
W-class fraction increases, whereas the top-12 Jaccard remains stable.

\begin{table*}[t]
\caption{\label{tab:S3_w}
W-class off-manifold stress-test metrics (10 seeds, 1000-shot).
}
\begin{ruledtabular}
\begin{tabular}{ccccc}
W fraction & Mixed fidelity & GHZ-subset fidelity & Top-12 Jaccard\\
\hline
0.00 & $0.9984 \pm 0.0003$ & $0.9982 \pm 0.0003$ & $1.000 \pm 0.000$\\
0.05 & $0.9484 \pm 0.0002$ & $0.9981 \pm 0.0002$ & $1.000 \pm 0.000$\\
0.10 & $0.8986 \pm 0.0002$ & $0.9980 \pm 0.0002$ & $1.000 \pm 0.000$\\
\end{tabular}
\end{ruledtabular}
\end{table*}

\section{KAN--MLP architecture-level comparison}\label{app:kan-mlp-comparison}

This note gives the quantitative details behind the width-matched
KAN--MLP comparison in Fig.~7 of the main text. The comparison is used as
architecture-level context, not as an exhaustive neural-architecture
benchmark.

\subsection{MLP internal support score definition}

The MLP does not natively provide pruned input-hidden-output pathway
objects. We therefore use a post-hoc internal support score,
$\mathrm{score}^{\mathrm{MLP}}_{i,o} = \sum_h |W^{(1)}_{i,h}|\,|W^{(2)}_{h,o}|$,
where $W^{(1)}$ and $W^{(2)}$ are the weight matrices of the first and
second MLP layers, respectively. This proxy is analogous to the
aggregated KAN pathway score defined in Sec.~II.D of the main text. It is
used only to enable a comparable top-channel ranking.

\subsection{KAN and MLP external-ablation channel rankings}

Table~\ref{tab:S4_ablation} reports the external-ablation top-12 channel rankings and
top-12 extended-channel Jaccard values for each model at the 1000-shot
condition.

\begin{table}[H]
\caption{\label{tab:S4_ablation}
KAN and MLP external-ablation channel rankings.
}
\begin{ruledtabular}
\begin{tabular}{lcc}
Metric & KAN & MLP\\
\hline
Top-12 extended Jaccard & $1.000 \pm 0.000$ & $0.601 \pm 0.048$\\
Mean $\Delta$RMSE (top-12) & $0.0126 \pm 0.0003$ & $0.0052 \pm 0.0011$\\
Clean-target fidelity & $0.9982 \pm 0.0003$ & $0.9856 \pm 0.0360$\\
\end{tabular}
\end{ruledtabular}
\end{table}

\subsection{Hidden-representation diagnostics}

Table~\ref{tab:S4_hidden} reports hidden-correlation and probe metrics for both
architectures.

\begin{table}[H]
\caption{\label{tab:S4_hidden}
KAN and MLP hidden-representation diagnostics (10 seeds, 1000-shot).
}
\begin{ruledtabular}
\begin{tabular}{lcc}
Metric & KAN & MLP\\
\hline
$\max|\mathrm{corr}(h,z)|$ & $0.972 \pm 0.037$ & $0.883 \pm 0.091$\\
$\max|\mathrm{corr}(h,c)|$ & $0.790 \pm 0.201$ & $0.766 \pm 0.171$\\
$\max|\mathrm{corr}(h,s)|$ & $0.885 \pm 0.149$ & $0.715 \pm 0.244$\\
Mean max $|\mathrm{corr}|$ & $0.882 \pm 0.057$ & $0.788 \pm 0.127$\\
Probe $R^2$ $(z,c,s)$ & $0.9985 \pm 0.0007$ & $0.9752 \pm 0.0728$\\
Probe $R^2$ $(c,s)$ & $0.9984 \pm 0.0007$ & $0.9457 \pm 0.1602$\\
\end{tabular}
\end{ruledtabular}
\end{table}

\subsection{Control-analysis checklist}

Table~\ref{tab:S3} provides the control-analysis checklist referenced in the main
text.

\begin{table*}[t]
\caption{\label{tab:S3}
Control-analysis checklist. For each control experiment, the table lists
the artifact tested, the evaluation metric, and the qualitative outcome.
}
\begingroup
\small
\setlength{\tabcolsep}{4pt}
\begin{ruledtabular}
\begin{tabular*}{\textwidth}{@{\extracolsep{\fill}}llll@{}}
Control & Artifact tested & Metric & Pass?\\
\hline
Shuffled input order & Column-position artifact & Top-12 Jaccard $=1.000$ & Yes\\
Decoy channels & Irrelevant-channel artifact & Decoy fraction $=0.000$ & Yes\\
Random-label & Dataset shortcut & Jaccard $=0.026$; $R^2 = 0.003$ & Yes\\
Noise-level stability & Noise fragility & Jaccard $=1.000$ (all levels) & Yes\\
Pruning-threshold & Sparsity artifact & Jaccard $\uparrow$ with sparsity & Yes\\
Selected-channel retraining & Attribution not functional & Fidelity $0.998$ (top-12) & Yes\\
90/10 training contamination & Manifold overfitting & Fidelity unchanged & Yes\\
W-class off-manifold & GHZ-manifold overfitting & Fidelity $\downarrow$, Jaccard stable & Boundary\\
10-seed extension & Initialization fragility & Jaccard $=1.000$ (all seeds) & Yes\\
20-seed hidden alignment & Hidden-space instability & $R^2 = 0.9985 \pm 0.0007$ & Yes\\
\end{tabular*}
\end{ruledtabular}
\endgroup
\end{table*}

\section{Pathway and edge-function audit trail}\label{app:pathway-audit}

This note records the audit trail for the pathway-level interpretation.
Pathway scores and top-$k$ fractions are defined in Sec.~II.D of the main
text. These diagnostics are used to inspect the dominant routes through
the trained KAN. They are not treated as an independent proof of perfect
pathway separation, because lower-ranked routes can include cross-output
or shared-support contributions.

\subsection{Dominant-pathway tables}

Tables~\ref{tab:S5_z}--\ref{tab:S5_s} list the highest-scoring pathways for each output
variable in the representative seed-0 1000-shot KAN. Each row gives the
input Pauli channel, mediating hidden unit, output variable, and pathway
score $\mathrm{score}_{i,o} = |\mathrm{scale}_{i,h}|
\,|\mathrm{scale}_{h,o}|$.

\begin{table}[H]
\caption{\label{tab:S5_z}
Dominant pathways for output $z$ (population imbalance).
Canonical support: $\{\mathrm{ZII},\mathrm{IZI},\mathrm{IIZ},\mathrm{ZZZ}\}$.
}
\begin{ruledtabular}
\begin{tabular}{clclc}
Rank & Input & Hidden unit & Output & Score\\
\hline
1 & ZZZ & $h_0$ & $z$ & 0.083\\
2 & IIZ & $h_0$ & $z$ & 0.071\\
3 & IZI & $h_1$ & $z$ & 0.067\\
4 & IIZ & $h_3$ & $z$ & 0.062\\
5 & ZII & $h_1$ & $z$ & 0.060\\
6 & IZI & $h_3$ & $z$ & 0.056\\
7 & ZII & $h_0$ & $z$ & 0.055\\
8 & ZZZ & $h_3$ & $z$ & 0.043\\
\end{tabular}
\end{ruledtabular}
\end{table}

\begin{table}[H]
\caption{\label{tab:S5_c}
Dominant pathways for output $c$ (real off-diagonal).
Canonical support: $\{\mathrm{XXX},\mathrm{XYY},\mathrm{YXY},\mathrm{YYX}\}$.
}
\begin{ruledtabular}
\begin{tabular}{clclc}
Rank & Input & Hidden unit & Output & Score\\
\hline
1 & YYX & $h_4$ & $c$ & 0.148\\
2 & XYY & $h_5$ & $c$ & 0.147\\
3 & YXY & $h_5$ & $c$ & 0.129\\
4 & XXX & $h_5$ & $c$ & 0.110\\
5 & ZZZ & $h_4$ & $c$ & 0.068\\
6 & ZII & $h_4$ & $c$ & 0.066\\
7 & IZI & $h_4$ & $c$ & 0.064\\
8 & XXX & $h_4$ & $c$ & 0.048\\
\end{tabular}
\end{ruledtabular}
\end{table}

\begin{table}[H]
\caption{\label{tab:S5_s}
Dominant pathways for output $s$ (imaginary off-diagonal).
Canonical support: $\{\mathrm{YXX},\mathrm{XYX},\mathrm{XXY},\mathrm{YYY}\}$.
}
\begin{ruledtabular}
\begin{tabular}{clclc}
Rank & Input & Hidden unit & Output & Score\\
\hline
1 & XXY & $h_2$ & $s$ & 0.172\\
2 & YYY & $h_2$ & $s$ & 0.167\\
3 & YXX & $h_2$ & $s$ & 0.165\\
4 & XYX & $h_2$ & $s$ & 0.158\\
5 & ZZZ & $h_0$ & $s$ & 0.024\\
6 & IZI & $h_1$ & $s$ & 0.023\\
7 & IIZ & $h_0$ & $s$ & 0.021\\
8 & ZII & $h_1$ & $s$ & 0.020\\
\end{tabular}
\end{ruledtabular}
\end{table}

\subsection{Aggregated pathway scores}

Aggregated pathway scores are computed as
$\mathrm{score}_{i,o} = \sum_h |\mathrm{scale}_{i,h}|\,|\mathrm{scale}_{h,o}|$
for each input--output pair, following Fig.~3 of the main text. The
row-level dominant pathways in Tables~\ref{tab:S5_z}--\ref{tab:S5_s} provide the corresponding
audit trail for the representative seed-0 1000-shot model.

\subsection{Top-$k$ pathway-support fractions}

Table~\ref{tab:S5_topk} reports the top-$k$ target-support fractions for each output.
The metric is defined in Sec.~II.D of the main text. The $k=4$ rows show
that the strongest pathways for each output lie on the expected canonical
support. The decrease at larger $k$, especially for the $c$ and $s$
outputs, indicates that lower-ranked pathways include additional shared or
cross-output channels. We therefore use this table to support the
dominant-pathway interpretation, not a claim that all retained pathways are
target-specific.

\begin{table}[H]
\caption{\label{tab:S5_topk}
Top-$k$ pathway-support fractions per output.
}
\begin{ruledtabular}
\begin{tabular}{cccc}
$k$ & $z$ & $c$ & $s$\\
\hline
4  & 1.000 & 1.000 & 1.000\\
8  & 1.000 & 0.625 & 0.500\\
12 & 0.917 & 0.500 & 0.333\\
\end{tabular}
\end{ruledtabular}
\end{table}

\subsection{Input-index mapping diagnostics}

The post-pruning input-index mapping check verifies that retained input
indices in the pruned KAN graph map back to the original 63 Pauli labels
(Table~\ref{tab:S1}). No index remapping errors were detected in any seed.

\subsection{Edge-function plots}

\begin{figure*}[t]
\centering
\includegraphics[width=0.85\textwidth]{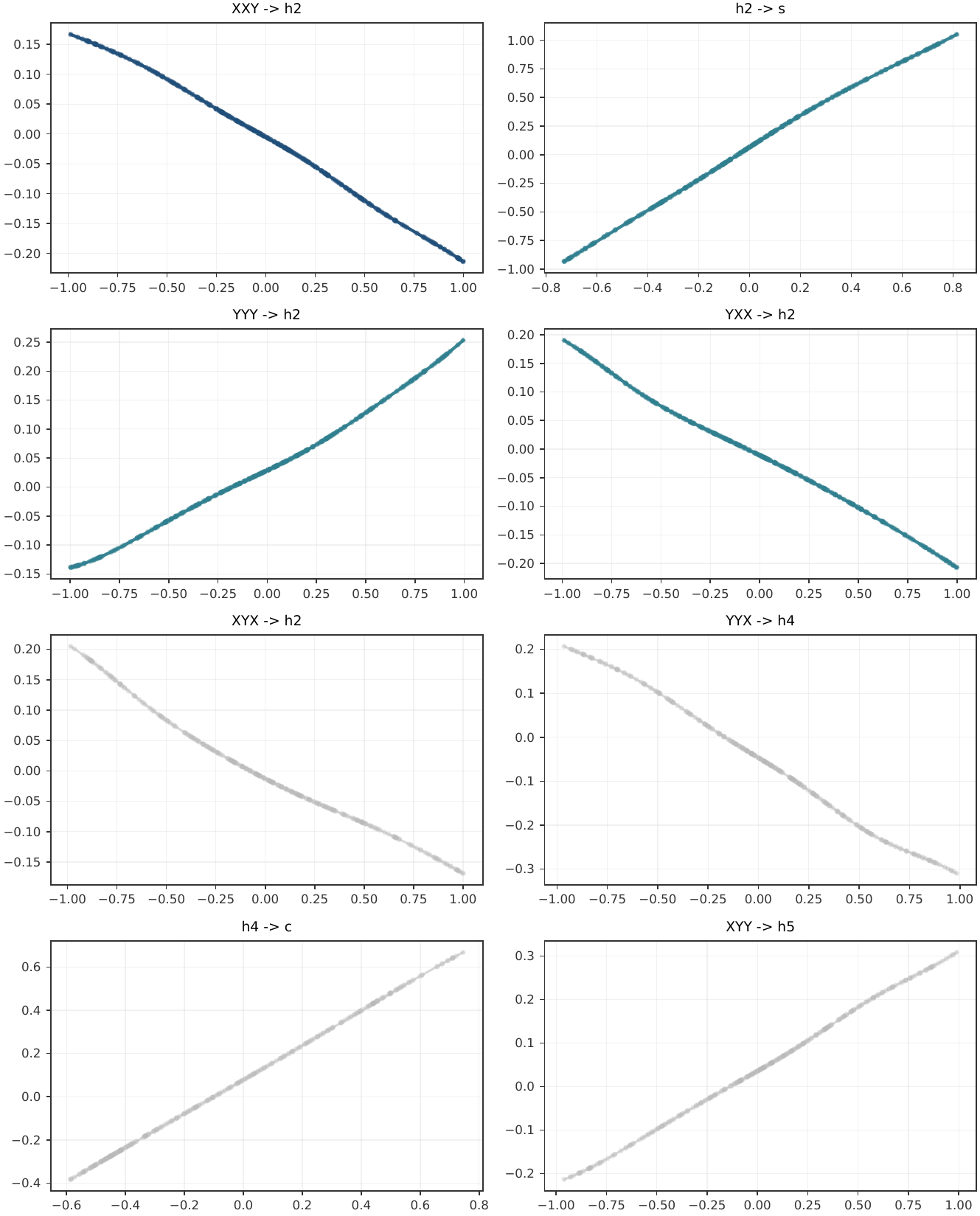}
\caption{\label{fig:S2}
Representative edge-function plots for high-importance edges in the
pruned KAN. Most functions are monotone and approximately linear, as
expected for this GHZ-family benchmark.
}
\end{figure*}

Figure~\ref{fig:S2} shows representative edge-function plots for high-importance
edges in the pruned KAN. The plotted functions are mostly monotone and
close to linear over the sampled range, which is consistent with the
nearly linear dependence of the GHZ target variables on the relevant Pauli
expectations. Thus, the figure is used as an audit trail for signs,
directions, and edge-level inspectability, rather than as evidence that
the benchmark requires richly nonlinear edge functions.

\section{Sparse formula recovery and linear baselines}\label{app:sparse-formula-linear-baselines}

This note gives the formula-level and linear-baseline results supporting
Tables~I and~II of the main text.

\subsection{Sparse formula recovery details}

Formula recovery follows the protocol described in Sec.~II.D of the main
text. Table~\ref{tab:S6_formula} reports formulas recovered from KAN outputs alongside the
true-variable baseline. All coefficients are given in Pauli-channel
units.

\begin{table*}[t]
\caption{\label{tab:S6_formula}
KAN-output and true-variable recovered formulas.
}
\begin{ruledtabular}
\begin{tabular}{llc}
Target & Recovered formula & $R^2$\\
\hline
$z$ (KAN)   & $0.2500\,(\mathrm{ZII}+\mathrm{IZI}+\mathrm{IIZ}+\mathrm{ZZZ})$ & 1.00000\\
$z$ (True)  & $0.2500\,(\mathrm{ZII}+\mathrm{IZI}+\mathrm{IIZ}+\mathrm{ZZZ})$ & 1.00000\\
$c$ (KAN)   & $0.2500\,(\mathrm{XXX}-\mathrm{XYY}-\mathrm{YXY}-\mathrm{YYX})$ & 1.00000\\
$c$ (True)  & $0.2500\,(\mathrm{XXX}-\mathrm{XYY}-\mathrm{YXY}-\mathrm{YYX})$ & 1.00000\\
$s$ (KAN)   & $0.2500\,(-\mathrm{YXX}-\mathrm{XYX}-\mathrm{XXY}+\mathrm{YYY})$ & 1.00000\\
$s$ (True)  & $0.2500\,(-\mathrm{YXX}-\mathrm{XYX}-\mathrm{XXY}+\mathrm{YYY})$ & 1.00000\\
\end{tabular}
\end{ruledtabular}
\end{table*}

\subsection{Full linear-baseline results}

Tables~\ref{tab:S6_linear_clean}--\ref{tab:S6_linear_depol} report OLS, ridge, LASSO, and elastic-net results across
three settings: clean data, 1000-shot finite-shot noise, and
depolarizing noise with $p=0.05$. Metrics are computed on the test set
and averaged over 3 seeds.

\begin{table*}[t]
\caption{\label{tab:S6_linear_clean}
Linear baseline results: clean (noiseless) GHZ-family data.
}
\begin{ruledtabular}
\begin{tabular}{lcccccc}
Model & Fidelity & RMSE & Top-12 Jacc. & Prec. & Rec. & Support\\
\hline
OLS   & $1.000 \pm 0.000$ & $\sim 10^{-16}$ & $1.000 \pm 0.000$ & 1.000 & 1.000 & 12/12\\
Ridge & $1.000 \pm 0.000$ & $\sim 10^{-16}$ & $1.000 \pm 0.000$ & 1.000 & 1.000 & 12/12\\
LASSO & $1.000 \pm 0.000$ & $\sim 10^{-9}$  & $0.528 \pm 0.157$ & 1.000 & 0.250 & 3/12\\
EN    & $1.000 \pm 0.000$ & $\sim 10^{-9}$  & $1.000 \pm 0.000$ & 1.000 & 0.583 & 7/12\\
\end{tabular}
\end{ruledtabular}
\end{table*}

\begin{table*}[t]
\caption{\label{tab:S6_linear_shot}
Linear baseline results: 1000-shot finite-shot noise.
}
\begin{ruledtabular}
\begin{tabular}{lcccccc}
Model & Fidelity & RMSE & Top-12 Jacc. & Prec. & Rec. & Support\\
\hline
OLS   & $0.9981 \pm 0.0001$ & $0.01263 \pm 0.00031$ & $1.000 \pm 0.000$ & 0.071 & 1.000 & 60/12\\
Ridge & $0.9979 \pm 0.0001$ & $0.01267 \pm 0.00031$ & $1.000 \pm 0.000$ & 0.074 & 1.000 & 60/12\\
LASSO & $0.9963 \pm 0.0003$ & $0.01318 \pm 0.00030$ & $1.000 \pm 0.000$ & 1.000 & 1.000 & 12/12\\
EN    & $0.9962 \pm 0.0003$ & $0.01322 \pm 0.00031$ & $1.000 \pm 0.000$ & 1.000 & 1.000 & 12/12\\
\end{tabular}
\end{ruledtabular}
\end{table*}

\begin{table*}[t]
\caption{\label{tab:S6_linear_depol}
Linear baseline results: depolarizing noise ($p = 0.05$). Fidelity is
clean-target fidelity.
}
\begin{ruledtabular}
\begin{tabular}{lcccccc}
Model & Clean fid. & RMSE & Top-12 Jacc. & Prec. & Rec. & Support\\
\hline
OLS   & $0.9750 \pm 0.0000$ & $0.02850 \pm 0.00008$ & $1.000 \pm 0.000$ & 1.000 & 1.000 & 12/12\\
Ridge & $0.9750 \pm 0.0000$ & $0.02850 \pm 0.00008$ & $1.000 \pm 0.000$ & 1.000 & 1.000 & 12/12\\
LASSO & $0.9750 \pm 0.0000$ & $0.02850 \pm 0.00008$ & $0.472 \pm 0.171$  & 1.000 & 0.250 & 3/12\\
EN    & $0.9750 \pm 0.0000$ & $0.02850 \pm 0.00008$ & $1.000 \pm 0.000$ & 1.000 & 0.583 & 7/12\\
\end{tabular}
\end{ruledtabular}
\end{table*}

\subsection{Support-identifiability caveat}

Under noiseless and purely depolarized conditions, LASSO selects a
single representative from a redundant GHZ-equivalent channel family
(e.g., $z \approx \mathrm{ZII}$) rather than the symmetric four-term
canonical formula. This behavior reflects collinearity among several
Pauli channels on the pure GHZ manifold. For example,
$\mathrm{ZII} = \mathrm{IZI} = \mathrm{IIZ} = \mathrm{ZZZ}$ when
evaluated on GHZ-family states alone. Finite-shot noise breaks this exact
degeneracy and allows LASSO and elastic net to recover the full four-term
canonical support. For the same reason, OLS and ridge can show perfect
support precision in the clean and purely depolarized tables, whereas
finite-shot noise populates otherwise zero channels and makes their
coefficient supports dense. Sec.~III.D of the main text discusses this caveat.

\subsection{Per-seed linear-baseline metrics}

Table~\ref{tab:S6_perseed} reports per-seed LASSO and elastic-net metrics at the
1000-shot condition, complementing the aggregate values in Table~II of
the main text.

\begin{table}[H]
\caption{\label{tab:S6_perseed}
Per-seed LASSO and elastic-net metrics (1000-shot, 3 seeds).
}
\begin{ruledtabular}
\begin{tabular}{llcccc}
Model & Seed & Fidelity & Jacc. & Prec. & Rec.\\
\hline
LASSO & 0 & 0.99655 & 1.000 & 1.000 & 1.000\\
LASSO & 1 & 0.99618 & 1.000 & 1.000 & 1.000\\
LASSO & 2 & 0.99603 & 1.000 & 1.000 & 1.000\\
Elastic net & 0 & 0.99647 & 1.000 & 1.000 & 1.000\\
Elastic net & 1 & 0.99610 & 1.000 & 1.000 & 1.000\\
Elastic net & 2 & 0.99583 & 1.000 & 1.000 & 1.000\\
\end{tabular}
\end{ruledtabular}
\end{table}

\end{document}